\documentclass[aps,prl,reprint,preprintnumbers,superscriptaddress,nofootinbib,longbibliography,floatfix]{revtex4-2}
\usepackage[T1]{fontenc}
\usepackage{lmodern}
\usepackage{amsmath,amssymb,graphicx,booktabs,bm,slashed}
\usepackage{microtype}
\usepackage{placeins}
\usepackage{xcolor}
\usepackage[hidelinks]{hyperref}
\hypersetup{pdftitle={Entanglement Bounds on Cardy Growth in Higher Dimensions},pdfauthor={Yuya Kusuki and Yin Tang}}
\newcommand{\dd}{\mathrm{d}}
\newcommand{\Tr}{\operatorname{Tr}}
\newcommand{\tr}{\operatorname{tr}}

\newcommand{\fs}{f_{\mathrm{s}}}
\newcommand{\Fs}{F_{\mathrm{s}}}
\newcommand{\as}{a_{\mathrm{s}}}
\newcommand{\fv}{f_{\mathrm{vac}}}
\newcommand{\ct}{C_T}
\newcommand{\cts}{C_{T,\mathrm{s}}}
\begin{document}
\preprint{KYUSHU-HET-379}
\preprint{RIKEN-iTHEMS-Report-26}
\title{Entanglement Bounds on Cardy Growth in Higher Dimensions}
\author{Yuya Kusuki}
\affiliation{\it Institute for Advanced Study, Kyushu University, Fukuoka 812-8581, Japan}
\affiliation{\it Department of Physics, Kyushu University, Fukuoka 819-0395, Japan}
\affiliation{\it RIKEN Center for Interdisciplinary Theoretical and Mathematical Sciences (iTHEMS), Wako, Saitama 351-0198, Japan}
\affiliation{\it JST, PRESTO, Kawaguchi, Saitama 332-0012, Japan}
\author{Yin Tang}
\affiliation{\it Institute for Advanced Study, Kyushu University, Fukuoka 812-8581, Japan}
\begin{abstract}
We propose bounds relating thermal entropy to entanglement in unitary conformal field theories: $f\le4\pi^2a$ in four dimensions and $f\le2.99830949\ldots\,F$ in three dimensions. Here $f$ is the dimensionless coefficient of the cosmological constant term in the thermal effective action, $a$ is the Euler anomaly coefficient, and $F$ is the sphere free energy. These bounds arise naturally as the high-temperature limits of proposed inequalities for spherical R\'enyi entropy that extend the conformal collider bound in four dimensions and an analogous conjecture in three dimensions. To verify this proposal, we formulate a stronger conjectural inequality common to both dimensions, expressed in terms of the stress-tensor three-point function and saturated by every free theory. We establish this inequality in controlled regimes of weakly coupled gauge theories, large-$N$ vector models, and holographic CFTs. Finite-$N$ numerical results provide additional support. Our proposal suggests that the equality between the universal coefficients governing the Cardy formula and entanglement entropy in two-dimensional CFTs extends to higher dimensions in the form of a universal inequality.
\end{abstract}
\maketitle

\begin{table*}[t]
\centering
\renewcommand{\arraystretch}{1.12}
\setlength{\tabcolsep}{4pt}
\begin{tabular}{@{}p{0.22\textwidth}p{0.25\textwidth}p{0.16\textwidth}p{0.32\textwidth}@{}}
\toprule
\raggedright \textbf{Class} & \raggedright \textbf{Representative examples} & \raggedright \textbf{Regime} & \raggedright \textbf{Key result}\tabularnewline
\midrule
\raggedright Gauge theories ($d=4$) & \raggedright Banks--Zaks; $\mathcal N=4$ Yang--Mills & \raggedright Weak coupling & \raggedright The leading correction to $\fv-f$ is nonnegative; see Eq.~\eqref{eq:weakgap}.\tabularnewline
\addlinespace[6pt]
\raggedright Ungauged vector models ($d=3$) & \raggedright Critical $O(N)$; Gross--Neveu & \raggedright Leading order in $1/N$ & \raggedright Thermal masses reduce entropy relative to massless free fields; see Eq.~\eqref{eq:vectorbound}.\tabularnewline
\addlinespace[6pt]
\raggedright Chern--Simons vector models ($d=3$) & \raggedright Regular/critical bosons and fermions & \raggedright Planar limit; finite 't Hooft coupling & \raggedright Convexity bounds the entropy averaged over holonomy in terms of the coefficients of $\langle TTT\rangle$; see Eq.~\eqref{eq:csbound}.\tabularnewline
\addlinespace[6pt]
\raggedright Holographic CFTs ($d=3,4$) & \raggedright $\mathcal N=4$ Yang--Mills; ABJM & \raggedright Classical Einstein gravity & \raggedright The null energy condition bounds the entropy density at fixed temperature by the planar Schwarzschild--AdS value.\tabularnewline
\midrule
\multicolumn{2}{@{}l}{Ising and $O(2)$ CFTs ($d=3$)} & \raggedright Finite $N$ & \raggedright Bootstrap + Monte Carlo: $f/\fv\simeq0.85$\tabularnewline
\bottomrule
\end{tabular}
\caption{Evidence for $f\le\fv$. See~\cite{SM} for the derivations. The last row gives approximate numerical results from Eq.~\eqref{eq:numerical}.}
\label{tab:summary}
\end{table*}

\paragraph*{Introduction.---}How strongly can entanglement constrain thermodynamics? In two-dimensional conformal field theory (CFT), the central charge determines both the universal contribution to entanglement entropy and the leading thermal free energy, and hence the high-energy density of states through the Cardy formula~\cite{Cardy}. Above two dimensions, conformal symmetry no longer fixes the thermal coefficient in terms of entanglement entropy. Entanglement may nevertheless bound this coefficient and thereby constrain thermal entropy and the growth of the high-energy density of states.

We define the dimensionless thermal coefficient $f$ as the coefficient of the cosmological constant term in the thermal effective action~\cite{Benjamin,Advant}. On a background $\mathbb{S}^1_\beta\times\Sigma$,
\begin{equation}
 I_{\rm th}=-\frac{f}{\beta^{d-1}}\int_\Sigma\dd^{d-1}x\sqrt g+\cdots,
 \qquad s=dfT^{d-1}.
 \label{eq:eft}
\end{equation}
Here $\Sigma$ is the spatial manifold, $g$ is the determinant of its metric, $d$ is the spacetime dimension, $T=\beta^{-1}$ is the temperature, and $s$ is the flat-space entropy density. The same coefficient controls the leading high-energy density of states~\cite{Shaghoulian,Benjamin}.

We compare $f$ with the Euler anomaly coefficient $a$ in four dimensions and the sphere free energy $F=-\log|Z(\mathbb{S}^3)|$ in three dimensions, where $Z(\mathbb{S}^3)$ is the partition function on the unit sphere. These coefficients determine the universal contribution to the entanglement entropy of a ball of radius $R$ in the vacuum~\cite{CHM,KPS}:
\begin{equation}
 S^{\rm univ}=\begin{cases}-F,&d=3,\\-4a\log(R/\epsilon),&d=4,\end{cases}
 \label{eq:entanglement}
\end{equation}
with ultraviolet cutoff $\epsilon$. Spherical R\'enyi entropy, with R\'enyi index $q$, relates entanglement to thermodynamics: at $q=1$ it gives the entanglement entropy, while its $q\to0$ limit probes the thermal coefficient~\cite{HMSY,Kusuki}.

In four dimensions, the conformal collider bound $c/a\le3$, where $c$ is the Weyl anomaly coefficient, suggests an extension to thermodynamics. This bound follows from positivity of the energy flux and is saturated by a free real scalar~\cite{Hofman,ColliderProof}. The connection is provided by spherical R\'enyi entropy, whose universal term is $-4a_q\log(R/\epsilon)$, with $a_1=a$. To first order in $1-q>0$, $c/a\le3$ is equivalent to requiring that $a_q/a$ not exceed its free-scalar value~\cite{PerlmutterRenyi}. Extending this bound throughout $0<q\le1$ would give $f/a\le4\pi^2$ in the thermal limit.

Our main proposal for unitary CFTs is the pair of sharp bounds
\begin{equation}
 f\le\begin{cases}\displaystyle\frac{8\pi\zeta(3)}{2\pi^2\log2-3\zeta(3)}\,F,&d=3,\\[7pt]4\pi^2a,&d=4.\end{cases}
 \label{eq:headline}
\end{equation}
The three-dimensional coefficient is $2.99830949\ldots$; both bounds are saturated by a free real scalar.

Equation~\eqref{eq:headline} is already strict for free fermions and, in four dimensions, free vectors. To examine the effects of interactions more directly, we use the stronger inequality $f\le\fv$. Here $\fv$ is a linear combination of the coefficients of the stress-tensor three-point function and equals $f$ in every free CFT. Unlike $4\pi^2a-f$, the difference $\fv-f$ vanishes for any free field content, so its nonnegativity constrains the interaction corrections themselves. The same definition applies in three and four dimensions, so results in both dimensions provide complementary evidence for the same conjecture. This stronger inequality implies the four-dimensional bound in Eq.~\eqref{eq:headline}. In three dimensions, it implies the corresponding bound when combined with the Bueno--Casini--Lasso Andino--Moreno (BCLM) conjecture for $\ct/F$, where $\ct$ is the coefficient of the stress-tensor two-point function~\cite{Bueno}. For related studies of entanglement entropy and conformal bounds in five dimensions, see Refs.~\cite{Willmore2024,BuenoFiveDimensions}.

Earlier work compared thermal entropy with $\ct$ in holographic CFTs~\cite{KovtunRitz} and with spherical entanglement entropy in large-$N$ vector models~\cite{Grable}. A separate conjecture states that thermal degrees of freedom decrease from an asymptotically free ultraviolet theory to the infrared~\cite{ACS}. Here the comparison with $\fv$ relates the thermal coefficient to the stress-tensor correlation functions of the same CFT. We identify common mechanisms that establish this stronger inequality in distinct classes, and test it at finite coupling and finite $N$ (Table~\ref{tab:summary}).

\paragraph*{Motivation from R\'enyi entropy.---}Write the universal terms in spherical R\'enyi entropy as $S_q^{\rm univ}=-F_q$ in three dimensions and $S_q^{\rm univ}=-4a_q\log(R/\epsilon)$ in four dimensions. The conformal map to a thermal state on hyperbolic space gives~\cite{HMSY,Kusuki}
\begin{equation}
 \begin{aligned}
 F_1&=F,& \lim_{q\to0}q^2F_q&=\frac{f}{2\pi},\\
 a_1&=a,& \lim_{q\to0}q^3a_q&=\frac{f}{16\pi^2}.
 \end{aligned}
 \label{eq:renyiends}
\end{equation}
Let $a_{q,\mathrm{s}}$ be the R\'enyi coefficient for a free real scalar, with $a_{1,\mathrm{s}}=a_{\mathrm{s}}=1/360$. We compare $a_q/a$ with its free-scalar value by defining
\begin{equation}
 R_4(q):=\frac{a_q/a}{a_{q,\mathrm{s}}/a_{\mathrm{s}}}.
 \label{eq:renyiratio}
\end{equation}
By construction, $R_4(1)=1$. Applying the identity $a'_1=-c/2$~\cite{PerlmutterRenyi} to both the theory and the free real scalar, for which $c=3a$, gives
\begin{equation}
 R'_4(1)=-\frac{c}{2a}+\frac32=\frac{3a-c}{2a}\ge0.
 \label{eq:renyimotivation}
\end{equation}
The inequality follows from the conformal collider bound $c\le3a$~\cite{Hofman,ColliderProof}. Thus, when $c<3a$, lowering $q$ from one makes $R_4(q)$ initially smaller than one: the normalized R\'enyi coefficient is bounded above by its free-scalar value for $q<1$ sufficiently close to one.

Motivated by this result near $q=1$, we conjecture $R_4(q)\le1$ throughout $0<q\le1$. Applying the thermal limits in Eq.~\eqref{eq:renyiends} to the theory and the free scalar gives
\begin{equation}
 \lim_{q\to0}R_4(q)=\frac{f}{4\pi^2a}.
 \label{eq:renyiratio-thermal}
\end{equation}
The proposed bound $f\le4\pi^2a$ is therefore the thermal limit of this more general conjecture.

In three dimensions, define
\begin{equation}
 R_3(q):=\frac{F_q/F}{F_{q,\mathrm{s}}/\Fs},
 \qquad R_3(1)=1,
 \label{eq:renyiratio-three}
\end{equation}
where $F_{q,\mathrm{s}}$ is the corresponding coefficient for a free real scalar and $\Fs=F_{1,\mathrm{s}}$. The identity $F'_1=-\pi^4\ct/12$~\cite{PerlmutterRenyi} gives
\begin{equation}
 R'_3(1)=\frac{\pi^4}{12}\left(\frac{\cts}{\Fs}-\frac{\ct}{F}\right)\ge0,
 \label{eq:renyimotivation-three}
\end{equation}
where $\cts$ is the free-scalar value of $\ct$. The inequality follows from the BCLM conjecture~\cite{Bueno}, which plays the role of the four-dimensional collider bound. This motivates the analogous conjecture $R_3(q)\le1$ for $0<q\le1$. Its thermal limit,
\begin{equation}
 \lim_{q\to0}R_3(q)=\frac{f/F}{\fs/\Fs},
 \label{eq:renyiratio-three-thermal}
\end{equation}
gives $f/F\le\fs/\Fs$, where $\fs$ is the thermal coefficient of a free real scalar.

These inequalities are verified throughout $0<q\le1$ for four-dimensional free CFTs and for three- and four-dimensional holographic CFTs~\cite{SM}.

\paragraph*{A stronger thermal inequality.---}To define $\fv$, we expand the parity-even part of the stress-tensor three-point function in the standard free-field basis~\cite{Hofman,ColliderProof,ParityCollider} and weight each coefficient by the thermal coefficient of the corresponding free field:
\begin{equation}
 \begin{aligned}
 \langle TTT\rangle_{\rm even}&=\sum_i n_i\langle TTT\rangle_i^{\rm free},\\
 \fv&:=\sum_i n_i f_i^{\rm free}.
 \end{aligned}
 \label{eq:vacdef}
\end{equation}
The index $i$ labels a real scalar ($s$) and a Majorana fermion ($M$) in three dimensions, or a real scalar ($s$), a Weyl fermion ($W$), and a Maxwell field ($V$) in four dimensions. Here $f_i^{\rm free}$ is the thermal coefficient of the corresponding free field. The coefficients $n_i$ need not be integers; positivity of the energy flux implies $n_i\ge0$, including in parity-violating three-dimensional theories.

Equality for each free field uniquely fixes the weights in Eq.~\eqref{eq:vacdef}, so $\fv$ for an interacting CFT is determined by $\langle TTT\rangle$ alone, without calculating its thermal free energy. We conjecture that it also bounds the thermal coefficient in interacting CFTs:
\begin{equation}
 f\le\fv.
 \label{eq:strong}
\end{equation}

In four dimensions, the free-field normalizations give
\begin{align}
 \frac{\fv}{\pi^2}&=\frac{n_s}{90}+\frac{7n_W}{360}+\frac{n_V}{45},\nonumber\\
 a&=\frac{n_s}{360}+\frac{11n_W}{720}+\frac{31n_V}{180}.
 \label{eq:fourdata}
\end{align}
Consequently,
\begin{equation}
 4\pi^2a-\fv=\pi^2\left(\frac{n_W}{24}+\frac{2n_V}{3}\right)\ge0.
 \label{eq:fourchain}
\end{equation}
The bound in terms of $a$ therefore follows from Eq.~\eqref{eq:strong} and positivity of the energy flux.

The same decomposition gives $(4\pi^2/5)(2c-a)-\fv=\pi^2n_W/120\ge0$, with equality if and only if $n_W=0$. Explicit four-dimensional comparisons, including supersymmetric multiplets and higher-curvature gravitational models, are given in~\cite{SM}.

In three dimensions, the thermal coefficient and sphere free energy of a free real scalar are
\begin{equation}
 \fs=\frac{\zeta(3)}{2\pi},\qquad \Fs=\frac{\log2}{8}-\frac{3\zeta(3)}{16\pi^2}.
 \label{eq:scalar3d}
\end{equation}
We normalize $\ct$ so that a real scalar has $\cts=3/(32\pi^2)$; a Majorana fermion has the same value. Using Eq.~\eqref{eq:vacdef} and the Ward identity relating the stress-tensor two- and three-point functions gives
\begin{equation}
 \fv=\fs\left(n_s+\frac34n_M\right),\qquad
 \frac{\ct}{\cts}=n_s+n_M.
 \label{eq:threedata}
\end{equation}
The BCLM conjecture is $0\le\ct/F\le\cts/\Fs$~\cite{Bueno}. Combining it with Eq.~\eqref{eq:strong} gives
\begin{equation}
 \frac f{\fs}\le\frac{\fv}{\fs}\le\frac{\ct}{\cts}\le\frac F{\Fs}.
 \label{eq:threechain}
\end{equation}
Only the last step requires the BCLM conjecture.

\paragraph*{A positive correction at weak coupling.---}Consider a four-dimensional weakly coupled gauge--Yukawa fixed point with a semisimple non-Abelian gauge group. In Euclidean signature, we normalize the Yukawa interactions and stable quartic potential as
\begin{equation}
 \begin{aligned}
 \mathcal L_Y&=\frac12\phi_a\left(Y^a_{ij}\psi_i\psi_j+Y^{a*}_{ij}\bar\psi_i\bar\psi_j\right),\\
 V_4&=\frac1{4!}\lambda_{abcd}\phi_a\phi_b\phi_c\phi_d.
 \end{aligned}
 \label{eq:interactions}
\end{equation}
Here $\phi_a$ and $\psi_i$ are real scalars and Weyl fermions with canonical kinetic terms; $Y^a$ and $\lambda_{abcd}$ are their Yukawa and quartic couplings. With $h\ll1$ a small expansion parameter, we take gauge and Yukawa couplings to be of order $h$ and quartic couplings of order $h^2$. For each simple gauge factor $A$, let $g_A$ be its coupling, $d_A$ its dimension, $C_A$ its adjoint quadratic Casimir, and $T_{S,A}$ the total Dynkin index of the real scalars.

In Banks--Zaks theories and the Litim--Sannino model~\cite{LitimSannino}, the scalars are absent or gauge singlets, so $T_{S,A}=0$. In weakly coupled $\mathcal N\ge1$ superconformal gauge theories, including supersymmetric QCD near the upper edge of its conformal window and $\mathcal N=4$ Yang--Mills theory, supersymmetry and the vanishing gauge beta function give $T_{S,A}=6C_A+O(h^2)$. More generally, for fixed points with $T_{S,A}\le6C_A$ at leading order, the corrections in the symmetric thermal phase with vanishing scalar background obey~\cite{SM}
\begin{equation}
 \begin{split}
 \fv-f\ge{}&\frac1{768}\sum_A g_A^2d_A(42C_A+5T_{S,A})\\
 &+\frac{\sum_a\tr(Y^{a\dagger}Y^a)}{384}
 +\frac{\lambda_{aabb}}{1152}+o(h^2).
 \end{split}
 \label{eq:weakgap}
\end{equation}
The trace is over gauge and flavor indices, and repeated scalar indices are summed. At this order, the correction to $\fv$ is a linear combination of gauge and Yukawa invariants with universal numerical coefficients, which we constrain using supersymmetric examples and conformal collider bounds~\cite{SM}. Combining the resulting lower bound on this correction with the known two-loop correction to $f$ gives Eq.~\eqref{eq:weakgap}. Vacuum stability at leading order requires $V_4\ge0$. Gaussian averaging then gives $\lambda_{aabb}=8\langle V_4\rangle_{\mathrm G}\ge0$, where $\langle\cdot\rangle_{\mathrm G}$ denotes an average over a Gaussian distribution with zero mean and unit covariance. Thus every leading term on the right-hand side of Eq.~\eqref{eq:weakgap} is nonnegative.

\paragraph*{Thermal masses at large $N$.---}Consider three-dimensional ungauged conformal vector models whose interactions can be expressed using a fixed number of scalar auxiliary fields. The matter consists of $N$ copies of $n_b$ real scalars and $n_D$ two-component Dirac fermions, with $n_b$ and $n_D$ fixed as $N\to\infty$. We work at leading order in $N$ in a homogeneous equilibrium phase with nonnegative squared scalar masses and real fermion masses.

At fixed values of the dimensionful auxiliary fields, denoted collectively by $\sigma$, the matter fields are free with masses determined by $\sigma$. Let $\mathcal F(T,\sigma)$ be the free-energy density at leading order in $N$ and $\sigma_*(T)$ the equilibrium values of the auxiliary fields. The saddle-point equations $\partial_\sigma\mathcal F=0$ and the chain rule give
\begin{equation}
 \frac{\dd\mathcal F(T,\sigma_*(T))}{\dd T}=\partial_T\mathcal F+\partial_\sigma\mathcal F\cdot\frac{\dd\sigma_*}{\dd T}=\partial_T\mathcal F.
 \label{eq:vectorstationarity}
\end{equation}
All partial derivatives are evaluated at $\sigma=\sigma_*(T)$, with $\partial_T$ taken at fixed $\sigma$. Since the interaction potential and the zero-point energies have no explicit temperature dependence at fixed $\sigma$, the entropy density is a sum of free-field entropy densities~\cite{Moshe,DeWolfeRomatschke}:
\begin{equation}
 s(T)=N\left[\sum_{a=1}^{n_b}s_B(T,m_a)+\sum_{\alpha=1}^{n_D}s_D(T,m_\alpha)\right],
 \label{eq:vectorentropy}
\end{equation}
where $s_B$ and $s_D$ are the entropy densities of a free real scalar and a free two-component Dirac fermion, respectively. The masses $m_a$ and $m_\alpha$ are evaluated at the equilibrium saddle.

At fixed temperature, the entropy of each free mode decreases with its energy. Since increasing the magnitude of the mass raises the energy at every momentum, each entropy density is bounded by its massless value. Writing $f_D=3\fs/2$ for the thermal coefficient of one massless Dirac fermion, we have
\begin{equation}
 \begin{aligned}
 s_B(T,m)&\le s_B(T,0)=3\fs T^2,\\
 s_D(T,m)&\le s_D(T,0)=3f_D T^2.
 \end{aligned}
 \label{eq:vectormassbound}
\end{equation}
In the conformally invariant zero-temperature vacuum, the leading stress-tensor three-point function is that of $Nn_b$ massless real scalars and $Nn_D$ massless Dirac fermions. The definition of $\fv$ and the conformal relation $s=3fT^2$ therefore turn the entropy comparison into
\begin{equation}
 f\le N(n_b\fs+n_Df_D)=\fv.
 \label{eq:vectorbound}
\end{equation}
The proof does not require solving the gap equations or evaluating the free energy at the saddle.

This explains the bound in the critical $O(N)$ model and in stable homogeneous phases of conformal scalar--fermion and sextic vector theories~\cite{Sachdev,Romatschke,DeWolfeRomatschke,Grable,Kvedaraite}. The critical Gross--Neveu model saturates it at leading order because its thermal saddle has zero fermion mass~\cite{Moshe}. The entropy comparison and the derivation of $\fv$ are given in~\cite{SM}.

\paragraph*{Chern--Simons vector theories.---}Consider three-dimensional $U(N)$ Chern--Simons theory with one fundamental flavor at leading planar order. The 't Hooft coupling is $\lambda=N/\kappa$, where $\kappa$ is the renormalized Chern--Simons level; we set $\theta=\pi|\lambda|$. In a homogeneous equilibrium phase with nonnegative pole masses, stationarity reduces the entropy to contributions of free massive fields averaged over the thermal holonomy~\cite{AGGMY}. This holonomy is the Wilson line of the gauge field around the thermal circle; its eigenvalues give phases in the matter boundary conditions.

Flux quantization bounds the density of holonomy phases, normalized to have unit integral, by $1/(2\theta)$~\cite{JainHolonomy}. For a mode of fixed energy, the entropy contribution decreases with the absolute value of the holonomy phase. The average over holonomy is therefore maximized by the uniform distribution on $[-\theta,\theta]$. For a bosonic mode of energy $E$, we write its contribution to the entropy, averaged over this distribution, as $(\sin\theta/\theta)H(\cos\theta,y)$, where $y=E/T$. At fixed $\theta$, $H(\cos\theta,y)$ decreases with $y$. Since reducing the pole mass lowers the energy at fixed momentum, setting the mass to zero gives an upper bound on the entropy. At fixed $y$, $H(x,y)$ is convex for $-1\le x\le1$, so
\begin{equation}
 H(\cos\theta,y)\le\tfrac{1+\cos\theta}{2}H(1,y)+\tfrac{1-\cos\theta}{2}H(-1,y).
 \label{eq:csentropycomparison}
\end{equation}
The endpoint values $H(1,y)$ and $H(-1,y)$ are the entropies of free bosonic and fermionic modes, respectively. Summing over the $2N$ real components of the fundamental scalar, integrating over momentum, and using $s=3fT^2$ gives~\cite{SM}
\begin{equation}
 \begin{aligned}
 f&\le N\frac{\sin\theta}{\theta}\bigl[(1+\cos\theta)\fs+(1-\cos\theta)f_M\bigr]\\
 &=n_s\fs+n_Mf_M=\fv.
 \end{aligned}
 \label{eq:csbound}
\end{equation}
Here $f_M=3\fs/4$ is the thermal coefficient of a free Majorana fermion. The coefficients multiplying $\fs$ and $f_M$ are exactly $n_s$ and $n_M$, determined independently from $\langle TTT\rangle$~\cite{AGY,ParityCollider}. Fermionic matter interchanges the two weights. The result applies to both regular and critical theories, nonperturbatively in $\lambda$.

\paragraph*{Holographic CFTs.---}Consider a CFT dual to classical Einstein gravity coupled to neutral scalar fields, with bulk action
\begin{equation}
 \begin{split}
 I={}&\frac1{16\pi G_{d+1}}\int\dd^{d+1}x\sqrt{-g}\\
 &\times\left[\mathcal R-\frac12G_{IJ}(\phi)\partial_\mu\phi^I\partial^\mu\phi^J-V(\phi)\right].
 \end{split}
 \label{eq:gravityaction}
\end{equation}
Here $G_{d+1}$ is Newton's constant, $\mathcal R$ is the Ricci scalar, $G_{IJ}$ is a positive-definite scalar kinetic matrix, and $V$ is the scalar potential. A static, homogeneous and isotropic thermal phase is dual to a regular planar black brane with metric
\begin{equation}
 \dd s^2=e^{2A(r)}[-h(r)\dd t^2+\dd\bm x^2]+\frac{\dd r^2}{h(r)}.
 \label{eq:gravitymetric}
\end{equation}
Here $r$ is the radial coordinate, $A(r)$ determines the warp factor $e^{2A(r)}$, and $h(r)$ is the blackening function, which vanishes at the horizon $r=r_H$. The scalar fields depend only on $r$ and have vanishing boundary sources.

For this metric, the null energy condition implies $A''\le0$, with primes denoting derivatives with respect to $r$. The Einstein equations, together with the asymptotic AdS conditions $A'(\infty)=1/L$ and $h(\infty)=1$, then yield $h'(r_H)\ge d/L$~\cite{SM}, where $L$ is the AdS radius. The Hawking temperature and Bekenstein--Hawking entropy density therefore give
\begin{equation}
 \begin{aligned}
 \frac{s}{T^{d-1}}&=\frac1{4G_{d+1}}\left(\frac{4\pi}{h'(r_H)}\right)^{d-1}\\
 &\le\frac1{4G_{d+1}}\left(\frac{4\pi L}{d}\right)^{d-1}.
 \end{aligned}
 \label{eq:gravityentropy}
\end{equation}
This bounds $s$ at fixed $T$ by its planar Schwarzschild--AdS value for the same $L$ and $G_{d+1}$. Using $s=dfT^{d-1}$ and the known stress-tensor three-point function gives
\begin{equation}
 \frac f\fv\le\begin{cases}\displaystyle\frac{2\pi^4}{189\zeta(3)}=0.8575\ldots,&d=3,\\[5pt]\displaystyle\frac34,&d=4.
 \end{cases}
 \label{eq:einstein}
\end{equation}
Saturation of Eq.~\eqref{eq:einstein} requires constant scalars. Thus $f<\fv$ throughout this class, independently of the scalar potential or the explicit solution. The same argument gives $f/(4\pi^2a)\le1/8$ and $f/F\le8\pi/27$, establishing the three-dimensional bound in this class without invoking the BCLM conjecture.

\paragraph*{Numerical evidence at finite $N$.---}Numerical results for the three-dimensional Ising and $O(2)$ CFTs support the conjecture beyond the analytic limits above. Estimates of $\langle TTT\rangle$ from the conformal bootstrap~\cite{IsingTTT,O2TTT}, together with Monte Carlo estimates of $f$~\cite{Bulgarelli}, give
\begin{equation}
 \begin{aligned}
 \left(\frac f{\fv}\right)_{\!\rm Ising}&=0.8484(22),\\
 \left(\frac f{\fv}\right)_{\!O(2)}&\simeq0.847.
 \end{aligned}
 \label{eq:numerical}
\end{equation}
For the Ising CFT, the independent fuzzy-sphere estimate $F=0.0612(5)$~\cite{Hu} also gives $f/[(\fs/\Fs)F]=0.834(7)$. The numerical inputs are given in~\cite{SM}.

\paragraph*{Thermal entropy and Cardy growth.---}Through $s=dfT^{d-1}$, Eq.~\eqref{eq:headline} bounds thermal entropy in terms of $a$ and $F$. On the unit $\mathbb{S}^{d-1}$, let $\rho(\Delta)$ denote the coarse-grained density of states of scaling dimension $\Delta$, including descendants. For $d=4$, the higher-dimensional Cardy formula~\cite{Shaghoulian,Benjamin} and Eq.~\eqref{eq:headline} give, as $\Delta\to\infty$,
\begin{equation}
 \log\rho(\Delta)\sim\frac43(6\pi^2f)^{1/4}\Delta^{3/4}\lesssim\frac43(24\pi^4a)^{1/4}\Delta^{3/4}.
 \label{eq:states}
\end{equation}
Similarly, for $d=3$,
\begin{equation}
 \log\rho(\Delta)\sim3(\pi f)^{1/3}\Delta^{2/3}\lesssim3\left(\pi\frac{\fs}{\Fs}F\right)^{1/3}\Delta^{2/3}.
 \label{eq:states3d}
\end{equation}
In two dimensions, the central charge fixes the leading Cardy growth of the density of states. The proposed bounds extend this connection to higher dimensions, where the universal contribution to spherical entanglement entropy would set an upper limit on that growth.

\paragraph*{Discussion.---}The analytic support for the proposed bounds comes from mechanisms that control whole classes of theories: positivity at weak coupling, the suppression of entropy by thermal masses and holonomy, and the null energy condition in gravity. Spherical R\'enyi entropy connects the four-dimensional proposal to the established inequality $c\le3a$, suggesting a route to a general proof beyond these controlled limits.

\begin{acknowledgments}
\textit{Acknowledgement.---} YK is supported by the INAMORI Frontier Program at Kyushu University, JSPS KAKENHI Grant Number 23K20046, MEXT Supporting Pioneering Research through AI for 1,000 Discovery challenges Program (SPReAD) Japan Grant Number JPMXP1726262823, and JST PRESTO Grant Number JPMJPR26A6. We used ChatGPT (OpenAI, GPT-6 astra) to assist with drafting and checking the manuscript under our direction. The authors are responsible for all content and conclusions.
\end{acknowledgments}

\paragraph*{Note added.---}While this manuscript was in preparation, Herzog and Stergiou~\cite{HerzogStergiou} reported overlapping conjectures and tests of thermal bounds in three-dimensional CFTs, including the bound on $f/F$. Our work was developed independently. Reference~\cite{HerzogStergiou} focuses on the free-scalar upper bound on $f/\ct$. Here we conjecture the stronger bound $f\le\fv$, formulated in terms of the stress-tensor three-point function in both three and four dimensions, and verify it in controlled regimes. Its saturation by every free CFT makes it a natural refinement of bounds based on entanglement. It yields $f\le4\pi^2a$ in four dimensions and $f/F\le\fs/\Fs$ in three dimensions as corollaries.
The R\'enyi generalizations used to motivate the entanglement bound also differ.

\FloatBarrier

\onecolumngrid
\clearpage
\setcounter{equation}{0}
\setcounter{section}{0}
\setcounter{subsection}{0}
\setcounter{table}{0}
\setcounter{figure}{0}
\setcounter{secnumdepth}{2}
\renewcommand{\theequation}{S\arabic{equation}}
\renewcommand{\thesection}{S\arabic{section}}
\renewcommand{\thetable}{S\arabic{table}}
\renewcommand{\thefigure}{S\arabic{figure}}
\renewcommand{\theHequation}{supplement.\arabic{equation}}
\renewcommand{\theHsection}{supplement.\arabic{section}}
\renewcommand{\theHtable}{supplement.\arabic{table}}
\renewcommand{\theHfigure}{supplement.\arabic{figure}}
\section*{Supplemental Material}
\label{sec:supplement}
\section{Definitions, normalizations, and the entanglement bounds}
\label{sec:SMconventions}

We consider unitary CFTs in spacetime dimension $d=3$ or $4$, and write $\beta=1/T$ for the inverse temperature. The coefficient $f$ of the cosmological constant term in the thermal effective action is equivalently defined by the flat-space partition function:
\begin{equation}
 \log Z(\mathbb{S}^1_\beta\times\mathbb R^{d-1})=fV\beta^{1-d},\qquad s=dfT^{d-1}.
 \label{eq:SMfdef}
\end{equation}
Here $V$ is the spatial volume and $s$ is the entropy density. We define the coefficient $C_T$ of the stress-tensor two-point function by
\begin{equation}
 \langle T_{\mu\nu}(x)T_{\rho\sigma}(0)\rangle=\frac{C_T}{|x|^{2d}}\mathcal I_{\mu\nu,\rho\sigma}(x),\qquad
 \mathcal I_{\mu\nu,\rho\sigma}=\frac12(I_{\mu\rho}I_{\nu\sigma}+I_{\mu\sigma}I_{\nu\rho})-\frac1d\delta_{\mu\nu}\delta_{\rho\sigma},
\end{equation}
where $I_{\mu\nu}=\delta_{\mu\nu}-2x_\mu x_\nu/x^2$. In four dimensions, let $a$ and $c$ denote the Euler and Weyl anomaly coefficients. Our conventions are
\begin{equation}
 C_T=\frac{40c}{\pi^4},\qquad
 \langle T^\mu{}_{\mu}\rangle=\frac{cW^2-aE_4}{16\pi^2},
 \label{eq:SManomaly}
\end{equation}
where $W^2$ is the squared Weyl tensor and $E_4$ is the Euler density; a scheme-dependent total derivative in the trace anomaly is omitted.

At separated points, the parity-even part of the stress-tensor three-point function decomposes into free-field structures. We define
\begin{equation}
 \langle TTT\rangle_{\rm even}=\sum_i n_i\langle TTT\rangle_i^{\rm free},\qquad
 f_{\rm vac}:=\sum_i n_i f_i^{\rm free}.
 \label{eq:SMvacdef}
\end{equation}
Here $i$ labels the free-field structures specified below, $n_i$ are their coefficients, and $f_i^{\rm free}$ is the thermal coefficient of the corresponding free field. The coefficients $n_i$ need not be integers. The conformal collider bounds imply $n_i\ge0$~\cite{Hofman,ColliderProof,ParityCollider}. In parity-violating three-dimensional theories, there is also a parity-odd tensor structure; positivity of the diagonal entries of the energy-flux matrix still implies $n_i\ge0$. We test the stronger inequality $f\le f_{\rm vac}$.

\subsection{Four dimensions}

With the tensor structures of one real scalar ($s$), one Weyl fermion ($W$), and one Maxwell field ($V$) as a basis, the free-field values give
\begin{align}
 f_{\rm vac}&=\pi^2\left(\frac{n_s}{90}+\frac{7n_W}{360}+\frac{n_V}{45}\right),\notag\\
 a&=\frac{n_s}{360}+\frac{11n_W}{720}+\frac{31n_V}{180},\qquad
 c=\frac{n_s}{120}+\frac{n_W}{40}+\frac{n_V}{10}.
 \label{eq:SMfourdata}
\end{align}
These expressions relate $f_{\rm vac}$ to the Euler anomaly:
\begin{equation}
 4\pi^2a-f_{\rm vac}=\pi^2\left(\frac{n_W}{24}+\frac{2n_V}{3}\right)\ge0.
 \label{eq:SMfourchain}
\end{equation}
The bound $f\le4\pi^2a$ therefore follows from $f\le f_{\rm vac}$ and the conformal collider bounds. The free real scalar has Euler anomaly $\as=1/360$. The quantity $f_{\rm vac}$ can also be expressed in terms of the conformal collider parameters $t_2,t_4$, which specify the angular dependence of the energy flux in a state created by the stress tensor~\cite{Hofman}:
\begin{equation}
 f_{\rm vac}=\pi^2c\left(\frac23+\frac{t_2}{9}+\frac{4t_4}{45}\right).
 \label{eq:SMcolliderfour}
\end{equation}

\paragraph*{Free supermultiplets.}Table~\ref{tab:SMfreefour} gives the values obtained by adding the free-field contributions in Eq.~\eqref{eq:SMfourdata}. These are free-theory values; they do not determine the thermal coefficient at a general point on an interacting conformal manifold.

\begin{table}[htbp]
\centering
\renewcommand{\arraystretch}{1.18}
\begin{tabular}{lccccc}
\toprule
Multiplet & $(n_s,n_W,n_V)$ & $a$ & $c$ & $f/\pi^2=f_{\rm vac}/\pi^2$ & $f/(4\pi^2a)$\\
\midrule
$\mathcal N=1$ chiral & $(2,1,0)$ & $1/48$ & $1/24$ & $1/24$ & $1/2$\\
$\mathcal N=1$ vector & $(0,1,1)$ & $3/16$ & $1/8$ & $1/24$ & $1/18$\\
$\mathcal N=2$ hypermultiplet & $(4,2,0)$ & $1/24$ & $1/12$ & $1/12$ & $1/2$\\
$\mathcal N=2$ vector & $(2,2,1)$ & $5/24$ & $1/6$ & $1/12$ & $1/10$\\
\bottomrule
\end{tabular}
\caption{Exact values for free four-dimensional supermultiplets, obtained by addition from Eq.~\eqref{eq:SMfourdata}. Entries are per component of the gauge representation; a full hypermultiplet is used. Every row satisfies $f=f_{\rm vac}$. Fermions obey thermal, antiperiodic boundary conditions.}
\label{tab:SMfreefour}
\end{table}

\paragraph*{Anomaly refinements.}The free-field decomposition gives a useful refinement involving only the anomalies:
\begin{equation}
 \frac{4\pi^2}{5}(2c-a)-f_{\rm vac}=\frac{\pi^2}{120}n_W\ge0.
 \label{eq:SMbosonicrefinement}
\end{equation}
Consequently, the stronger conjecture implies the hierarchy
\begin{equation}
 f\le f_{\rm vac}\le\frac{4\pi^2}{5}(2c-a)
 \le\frac{4\pi^2}{3}c\le4\pi^2a.
 \label{eq:SMfourhierarchy}
\end{equation}
The last two steps use $c\le3a$. The intermediate anomaly bound is saturated by free scalars and free Maxwell fields. For an $\mathcal N\ge1$ SCFT, the superconformal Ward identity in Eq.~\eqref{sm:weak:susyidentity} gives the stronger conjectured ceiling $f\le(2\pi^2/3)(2c-a)$. The identity is specific to superconformal theories. Already for free nonsupersymmetric fields,
\begin{equation}
 f_0-\frac{2\pi^2}{3}(2c_0-a_0)
 =\frac{\pi^2}{540}(n_s-2n_W+2n_V)
 \label{eq:SMnonsusywarning}
\end{equation}
can be positive, where the subscript $0$ denotes free-theory values.

\paragraph*{Non-Lagrangian targets with known vacuum data.}For the rank-one Argyres--Douglas theories $H_0,H_1,H_2$ and Minahan--Nemeschansky theories with $E_6,E_7,E_8$ flavor symmetry, the anomaly coefficients are known~\cite{AharonyTachikawa,ShapereTachikawa}. The superconformal identity in Eq.~\eqref{sm:weak:susyidentity} therefore determines $f_{\rm vac}$ exactly, as listed in Table~\ref{tab:SMnonlagrangian}. The numbers are proposed upper bounds on an independently measurable thermal coefficient, not computations of that coefficient.

\begin{table}[htbp]
\centering
\renewcommand{\arraystretch}{1.2}
\begin{tabular}{lcccc}
\toprule
Theory & $a$ & $c$ & $f_{\rm vac}/\pi^2=\frac23(2c-a)$ & $4a$\\
\midrule
$H_0=(A_1,A_2)$ & $43/120$ & $11/30$ & $1/4$ & $43/30$\\
$H_1=(A_1,A_3)$ & $11/24$ & $1/2$ & $13/36$ & $11/6$\\
$H_2=(A_1,D_4)$ & $7/12$ & $2/3$ & $1/2$ & $7/3$\\
$E_6$ & $41/24$ & $13/6$ & $7/4$ & $41/6$\\
$E_7$ & $59/24$ & $19/6$ & $31/12$ & $59/6$\\
$E_8$ & $95/24$ & $31/6$ & $17/4$ & $95/6$\\
\bottomrule
\end{tabular}
\caption{Exact vacuum data and conjectured thermal ceilings $f/\pi^2\le f_{\rm vac}/\pi^2\le4a$ for non-Lagrangian four-dimensional CFTs. No ordinary thermal $f$ is inferred from the anomalies. A supersymmetric index uses different boundary conditions and cannot replace the thermal partition function in this test.}
\label{tab:SMnonlagrangian}
\end{table}

\subsection{Three dimensions}

We define the sphere free energy by $F=-\log|Z(\mathbb{S}^3)|$, where $Z(\mathbb{S}^3)$ is the partition function on the unit sphere. For a free real scalar, the thermal coefficient $\fs$ and sphere free energy $\Fs$ are~\cite{KPS,KlebanovRenyi}
\begin{align}
 \fs&=\frac{\zeta(3)}{2\pi},& \Fs&=\frac{\log2}{8}-\frac{3\zeta(3)}{16\pi^2}.
 \label{eq:SMthreefree}
\end{align}
A free Majorana fermion ($M$) has $f_M=3\fs/4$. The free real scalar and free Majorana fermion have the same coefficient $\cts=3/(32\pi^2)$ in their stress-tensor two-point functions. For a general CFT, the Ward identity relating $\langle TTT\rangle$ to $\langle TT\rangle$ then gives $C_T=(n_s+n_M)\cts$, where $n_s$ and $n_M$ are the coefficients of the corresponding free-field structures in Eq.~\eqref{eq:SMvacdef}. In this basis, $f_{\rm vac}$ and the three-dimensional conformal collider parameter $t_4$ are
\begin{equation}
 f_{\rm vac}=\fs\left(n_s+\frac34n_M\right)
 =\fs\frac{C_T}{\cts}\frac{28+t_4}{32},\qquad
 t_4=4\frac{n_s-n_M}{n_s+n_M}.
 \label{eq:SMcolliderthree}
\end{equation}
Positivity of the energy flux gives $-4\le t_4\le4$ and therefore $f_{\rm vac}\le \fs C_T/\cts$. To relate this upper bound to $F$, we use the Bueno--Casini--Lasso Andino--Moreno (BCLM) conjecture~\cite{Bueno}:
\begin{equation}
 0\le\frac{C_T}{F}\le\frac{\cts}{\Fs}=\frac{3}{4\pi^2\log2-6\zeta(3)}.
\end{equation}
Combining the BCLM conjecture with $f\le f_{\rm vac}$ gives
\begin{equation}
 f\le f_{\rm vac}\le \fs\frac{C_T}{\cts}\le\frac{\fs}{\Fs}F,
 \qquad \frac{\fs}{\Fs}=2.998309492369\ldots.
 \label{eq:SMthreechain}
\end{equation}

For comparison, a free two-component Dirac fermion has
\begin{equation}
 f_D=\frac32\fs,\qquad F_D=\frac{\log2}{4}+\frac{3\zeta(3)}{8\pi^2}.
\end{equation}
A free $\mathcal N=2$ chiral multiplet (one complex scalar and one Dirac fermion) has $f=7\fs/2$, $F=\log2/2$, and therefore $f/F=7\zeta(3)/(2\pi\log2)=1.932\ldots$; in both examples $f=f_{\rm vac}$. A decoupled unitary topological sector contributes neither to $f$ nor to stress-tensor correlation functions at separated points. Its contribution to the sphere free energy is $F_{\rm top}=\log\mathcal D$, where $\mathcal D\ge1$ is the total quantum dimension. Such a sector leaves $f\le f_{\rm vac}$ unchanged and weakens the proposed upper bound in terms of $F$.

\section{Weakly coupled gauge theories in four dimensions}
\label{sm:weak:section}

We prove $f\le\fv$ at leading nontrivial order for weakly coupled fixed points obeying Eq.~\eqref{sm:weak:scope}. The proof uses the conformal collider bounds and supersymmetric Ward identities, but does not require supersymmetry of the theory under consideration.

\subsection{Conventions and result}

Consider a gauge theory with a semisimple non-Abelian gauge group $\prod_A G_A$ with couplings $g_A$, real scalars $\phi_a$, and Weyl fermions $\psi_i$. The gauge-invariant Euclidean action in flat space is $\int\dd^4x\,\mathcal L_E$, with
\begin{equation}
 \begin{aligned}
 \mathcal L_E={}&\frac14\sum_A F_{\mu\nu}^{A,I}F_{\mu\nu}^{A,I}+\frac12(D_\mu\phi)_a(D_\mu\phi)_a+\bar\psi_i\bar\sigma_E^\mu(D_\mu\psi)_i\\
 &+\frac12\phi_a\left(Y^a_{ij}\psi_i\psi_j+Y^{a*}_{ij}\bar\psi_i\bar\psi_j\right)+V_4(\phi),\qquad
 V_4(\phi)=\frac1{4!}\lambda_{abcd}\phi_a\phi_b\phi_c\phi_d.
 \end{aligned}
 \label{sm:weak:interactions}
\end{equation}
Here $\psi_i,\bar\psi_i$ are independent Grassmann fields and $\bar\sigma_E^\mu=(i\boldsymbol\sigma,\mathbf1)$, with Pauli matrices $\boldsymbol\sigma$. The covariant derivative and field strengths are
\begin{equation}
 \begin{aligned}
 D_\mu&=\partial_\mu-i\sum_A g_A A_\mu^{A,I}T_A^I,\\
 F_{\mu\nu}^{A,I}&=\partial_\mu A_\nu^{A,I}-\partial_\nu A_\mu^{A,I}+g_A f_A^{IJK}A_\mu^{A,J}A_\nu^{A,K}.
 \end{aligned}
 \label{sm:weak:covariant}
\end{equation}
Here $A_\mu^{A,I}$ is the gauge field for $G_A$, $T_A^I$ are Hermitian generators in the representation of the field on which $D_\mu$ acts, and $f_A^{IJK}$ are the structure constants. Repeated field and adjoint indices are summed. The fields are canonically normalized, so the gauge couplings enter through the covariant derivatives and the nonlinear terms in the field strengths. The matrices $Y^a$ are symmetric, and $\lambda_{abcd}$ is fully symmetric. Let $C_A=C_2(G_A)$ be the quadratic Casimir in the adjoint representation. The quantities $T_{F,A}$ and $T_{S,A}$ denote the sums of Dynkin indices of Weyl fermions and real scalars, including multiplicities from the other gauge factors. A complex scalar in a representation $R$ contributes $2T(R)$ to $T_{S,A}$, where $T(R)$ is its Dynkin index.

We consider weakly coupled fixed points near a Gaussian theory, with gauge and Yukawa couplings of order $h$ and quartic couplings of order $h^2$, where $h$ is a small expansion parameter. At the fixed point, each gauge beta function satisfies
\begin{equation}
 \beta_{g_A}=-\frac{g_A^3}{16\pi^2}\left[\frac{11}{3}C_A-\frac23T_{F,A}-\frac16T_{S,A}+O(h^2)\right]=0.
 \label{sm:weak:gaugebeta}
\end{equation}
For $g_A\neq0$, this implies
\begin{equation}
 4T_{F,A}+T_{S,A}=22C_A+O(h^2).
 \label{sm:weak:beta}
\end{equation}
With $d_A=\dim G_A$ and $\tr$ denoting the trace over gauge and flavor indices, define
\begin{equation}
 D=\sum_Ag_A^2d_AC_A,\qquad S=\sum_Ag_A^2d_AT_{S,A},\qquad
 Y^2=\sum_a\tr(Y^{a\dagger}Y^a).
 \label{sm:weak:invariants}
\end{equation}
In Banks--Zaks theories and the Litim--Sannino model~\cite{LitimSannino}, the scalars are absent or gauge singlets, so $S=0$. In weakly coupled $\mathcal N\ge1$ superconformal gauge theories, including supersymmetric QCD near the upper edge of its conformal window and $\mathcal N=4$ Yang--Mills theory, the supersymmetric field content gives $T_{S,A}=2(T_{F,A}-C_A)$. Together with Eq.~\eqref{sm:weak:beta}, this implies $S=6D+O(h^4)$. These examples therefore satisfy
\begin{equation}
 S\le6D+O(h^4).
 \label{sm:weak:scope}
\end{equation}
More generally, for weakly coupled fixed points obeying this relation, we will show that
\begin{equation}
 \fv-f\ge\frac{42D+5S}{768}+\frac{Y^2}{384}+\frac{\lambda_{aabb}}{1152}+o(h^2).
 \label{sm:weak:gap}
\end{equation}
The gauge and Yukawa contributions on the right-hand side are nonnegative. The sign of the quartic contribution follows from vacuum stability at leading order: since $V_4$ is homogeneous of degree four, boundedness from below requires $V_4(\phi)\ge0$ for all $\phi$. Gaussian averaging then gives $\lambda_{aabb}\ge0$, as shown in Eq.~\eqref{sm:weak:stability}. Thus every leading contribution in Eq.~\eqref{sm:weak:gap} is nonnegative.

\subsection{Stress-tensor correlation functions}

Let $f_0$ denote the thermal coefficient of the free theory with the same field content. At zero coupling, $f=\fv=f_0$, so verifying $f\le\fv$ amounts to comparing the corrections induced by interactions. We use $\delta$ to denote corrections to free-theory values and define
\begin{equation}
 \delta f=f-f_0,\qquad \delta\fv=\fv-f_0.
 \label{sm:weak:corrections}
\end{equation}
The desired inequality is thus $\delta f\le\delta\fv$. To constrain $\delta\fv$, we rewrite Eq.~\eqref{eq:SMfourdata} as
\begin{equation}
 \frac{\fv}{\pi^2}=\frac{13c}{12}-\frac a2+\frac{n_s}{288},
 \label{sm:weak:vacidentity}
\end{equation}
where $n_s$ is the coefficient of the free-scalar tensor structure in $\langle TTT\rangle$. At a perturbative fixed point, $\delta a=O(h^4)$~\cite{DondiProchazkaSannino}, so only $\delta c$ and $\delta n_s$ contribute to $\delta\fv$ at order $h^2$.

We next determine the possible dependence of $\delta\fv$ on the couplings at order $h^2$. The gauge contributions to stress-tensor correlators at separated points contain two generators and are linear combinations of $g_A^2d_AC_A$, $g_A^2d_AT_{F,A}$, and $g_A^2d_AT_{S,A}$. Different simple gauge factors cannot mix at this order. Two Yukawa vertices give the single invariant $Y^2$. Mixed gauge--Yukawa terms vanish because the scalar and gauge-field legs of one gauge and one Yukawa vertex cannot all be contracted.

Quartic couplings are also of order $h^2$, so we must check that they introduce no additional contribution. They can enter either through the stress tensor itself or through an insertion of $V_4$. The potential term in $T_{\mu\nu}$ is proportional to the metric and is removed by transverse-traceless projection, which also removes scalar improvement terms. For an insertion of $V_4$, massless tadpoles vanish in dimensional regularization. In the remaining connected diagrams, two legs of the quartic vertex must attach to the same bilinear stress-tensor insertion. Summing the corresponding contractions gives a factor of the free two-point function $\langle T_{\mu\nu}(x)(\phi_a\phi_b)(y)\rangle$, which vanishes at separated points. Thus quartic couplings do not contribute to $\delta\fv$ at this order. Contact terms do not affect the energy flux and hence do not change the coefficients entering $\fv$.

Eliminating $T_{F,A}$ with Eq.~\eqref{sm:weak:beta} therefore gives
\begin{equation}
 \delta\fv=\alpha D+\beta S+\gamma Y^2+O(h^4),
 \label{sm:weak:linearform}
\end{equation}
where $\alpha$, $\beta$, and $\gamma$ are universal numerical coefficients: at this order, the gauge and Yukawa factors are contained in $D$, $S$, and $Y^2$, leaving the same loop integrals for every theory. We determine $\gamma$ and $\alpha+6\beta$ using two supersymmetric examples with $S=6D$ at leading order but different values of $Y^2/D$. A Banks--Zaks theory then determines the sign of $\alpha$. For the supersymmetric examples, the $\mathcal N=1$ superconformal Ward identity $n_s=2(n_W-n_V)$~\cite{Hofman} expresses $\fv$ in terms of $a$ and $c$:
\begin{equation}
 \fv=\frac{2\pi^2}{3}(2c-a).
 \label{sm:weak:susyidentity}
\end{equation}
Supersymmetric QCD near the upper edge of its conformal window has $T_{S,A}=6C_A+O(h^2)$, so $S=6D+O(h^4)$. Its Yukawa interactions between matter and gauginos give $Y^2=2S=12D+O(h^4)$. The one-loop anomalous dimensions~\cite{MartinVaughn} and the superconformal anomaly formula~\cite{Anselmi} give
\begin{equation}
 \delta c=\frac{D}{64\pi^2}+O(h^4),\qquad
 \delta\fv=\frac D{48}+O(h^4).
 \label{sm:weak:sqcd}
\end{equation}
In $\mathcal N=4$ Yang--Mills theory, $S=6D$, $Y^2=24D$, and the protected stress-tensor three-point function gives $\delta\fv=0$. Matching the leading terms for these two theories gives
\begin{equation}
 \alpha+6\beta+12\gamma=\frac1{48},\qquad
 \alpha+6\beta+24\gamma=0.
 \label{sm:weak:matching}
\end{equation}
Subtracting these equations gives $\gamma=-1/576$, and hence $\alpha+6\beta=1/24$. To determine the sign of the remaining independent coefficient $\alpha$, consider a Banks--Zaks theory, for which $S=Y^2=0$ and hence $\delta\fv=\alpha D+O(h^4)$. In the Gaussian limit of this theory, $n_s=0$, so the conformal collider bound $n_s\ge0$ requires the leading correction $\delta n_s$ to be nonnegative~\cite{Hofman,ColliderProof}. At the fixed point, the known correction to $c$ is $\delta c=5D/(256\pi^2)+O(h^4)$~\cite{ZollerChetyrkin,OsbornStergiou}. Together with $\delta a=O(h^4)$, Eq.~\eqref{sm:weak:vacidentity} gives a positive leading correction to $\fv$. More precisely, comparing the order-$h^2$ terms gives
\begin{equation}
 \alpha D=\frac{65D}{3072}+\frac{\pi^2}{288}\delta n_s+O(h^4).
 \label{sm:weak:alphabound}
\end{equation}
Positivity of the leading correction to $n_s$ therefore implies $\alpha\ge65/3072>0$. Since $\alpha$ is universal, this constraint applies to all theories under consideration.

Eliminating $\beta$ and $\gamma$ using these relations gives
\begin{equation}
 \delta\fv+\frac{Y^2}{576}=\alpha\left(D-\frac S6\right)+\frac S{144}+O(h^4).
 \label{sm:weak:positiveform}
\end{equation}
The condition in Eq.~\eqref{sm:weak:scope} makes $D-S/6$ nonnegative at order $h^2$. Since $\alpha\ge0$ and $S\ge0$, both terms on the right-hand side are nonnegative at this order, yielding
\begin{equation}
 \delta\fv\ge-\frac{Y^2}{576}+O(h^4).
 \label{sm:weak:vacuum}
\end{equation}

\subsection{Thermal free energy and stability}

To compute $\delta f$, we work in the symmetric thermal phase with equilibrium scalar background $\phi_a=0$. This assumption is satisfied in the weakly coupled examples listed above. We use the known two-loop gauge and fermion contributions to the thermal free energy~\cite{ArnoldZhai}. At this order, the scalar--gauge and Yukawa terms depend only on $g_A^2d_AT_{S,A}$ and $Y^2$, respectively. The scalar--gauge and Yukawa coefficients follow from the Standard Model calculation~\cite{GyntherVepsalainen}, expressed in the conventions of Eq.~\eqref{sm:weak:interactions}. Together with the quartic term evaluated below, these contributions give
\begin{equation}
 -\delta f=\sum_A\frac{g_A^2d_A}{144}\left[C_A+\frac54(T_{F,A}+T_{S,A})\right]
 +\frac{5Y^2+\lambda_{aabb}}{1152}+o(h^2).
 \label{sm:weak:thermal}
\end{equation}
Bosonic screening contributes at order $h^3$, beyond the order displayed: the scalar and electric screening masses $m=O(hT)$ give zero-mode contributions of order $Tm^3=O(h^3T^4)$ to the free-energy density~\cite{BraatenNietoEFT,BraatenNietoQCD}. The quartic contribution follows from the free massless scalar propagator at coincident points. Subtracting its zero-temperature part gives
\begin{equation}
 \langle\phi_a(x)\phi_b(x)\rangle_T-\langle\phi_a(x)\phi_b(x)\rangle_{T=0}=\delta_{ab}\int\frac{\dd^3p}{(2\pi)^3}\frac{1}{p(e^{p/T}-1)}=\delta_{ab}\frac{T^2}{12},
 \label{sm:weak:scalarcontraction}
\end{equation}
where $p=|\mathbf p|$. In dimensional regularization, the zero-temperature contraction vanishes. The three Wick contractions of $V_4$ therefore contribute $\lambda_{aabb}(T^2/12)^2/8=\lambda_{aabb}T^4/1152$ to the free-energy density.

Since $V_4(\phi)\ge0$ at leading order, its average over a centered Gaussian distribution with unit covariance is also nonnegative. The three Wick contractions give
\begin{equation}
 \langle V_4\rangle_{\mathrm G}=\frac{3}{4!}\sum_{a,b}\lambda_{aabb}=\frac18\sum_{a,b}\lambda_{aabb}\ge0.
 \label{sm:weak:stability}
\end{equation}
This constrains the summed contraction even when individual quartic couplings are negative. Moreover, for a continuous nonnegative quartic polynomial the Gaussian average vanishes only if the polynomial is identically zero. Substituting Eq.~\eqref{sm:weak:beta} into Eq.~\eqref{sm:weak:thermal} and combining the result with Eq.~\eqref{sm:weak:vacuum} proves Eq.~\eqref{sm:weak:gap}.

The Banks--Zaks argument above gives the stronger bound $\alpha\ge65/3072$. Using this bound and retaining the positive term $S/144$ in Eq.~\eqref{sm:weak:positiveform} yields
\begin{equation}
 f_{\rm vac}-f\ge\frac{42D+5S}{768}+\frac{S}{144}
 +\frac{65}{3072}\left(D-\frac S6\right)
 +\frac{Y^2}{384}+\frac{\lambda_{aabb}}{1152}+o(h^2)
 \label{sm:weak:sharpergap}
\end{equation}
under the condition in Eq.~\eqref{sm:weak:scope}. The simpler bound in the main text follows by discarding the two additional nonnegative leading contributions. The higher-order corrections cannot reverse a strictly positive contribution at order $h^2$ for sufficiently small couplings.

\subsection{Application to \texorpdfstring{$\mathcal N=2$}{N=2} superconformal QCD}

For $\mathcal N=2$ $SU(N_c)$ SQCD with $N_f=2N_c$ full fundamental hypermultiplets, the protected anomaly data and the free-limit thermal coefficient are
\begin{equation}
 a=\frac{7N_c^2-5}{24},\qquad c=\frac{2N_c^2-1}{6},\qquad
 f_{\rm vac}=f_0=\frac{\pi^2(3N_c^2-1)}{12}.
 \label{eq:SMN2SQCD}
\end{equation}
Thus $f_0/(4\pi^2a)=(3N_c^2-1)/[2(7N_c^2-5)]\to3/14$ as $N_c\to\infty$. The anomaly coefficients, and hence $f_{\rm vac}$, are protected along the conformal manifold, whereas the thermal coefficient $f$ is not. Equation~\eqref{sm:weak:gap} implies $f<f_{\rm vac}$ at sufficiently small nonzero gauge coupling; $f<4\pi^2a$ already holds at the free limit and persists nearby. Determining the ordinary thermal coefficient beyond weak coupling would provide a further test of the bound.

\section{Thermal masses in three-dimensional vector models}
\label{sm:vec:section}

We consider three-dimensional conformal vector models whose interactions can be expressed in terms of scalar auxiliary fields $\sigma$~\cite{Moshe,DeWolfeRomatschke}. The matter content consists of $n_b$ species of real scalars $\phi_{ia}$ and $n_D$ species of two-component Dirac fermions $\psi_{i\alpha}$, where $i=1,\ldots,N$ is the vector index. We take the limit $N\to\infty$ with $n_b$, $n_D$, and the number of auxiliary fields fixed. At leading order in this limit, the entropy density is a sum of contributions from free fields with the corresponding thermal masses. Since each contribution decreases as the magnitude of the mass increases, we can prove $f\le\fv$ without solving the gap equations.

The Euclidean action at this order takes the form
\begin{equation}
 I_E=\int_0^\beta\dd\tau\int\dd^2x\,\left\{NU(\sigma)+\frac12\sum_{i=1}^N\sum_{a=1}^{n_b}\left[(\partial_\mu\phi_{ia})^2+m_a^2(\sigma)\phi_{ia}^2\right]+\sum_{i=1}^N\sum_{\alpha=1}^{n_D}\bar\psi_{i\alpha}\left[\gamma^\mu\partial_\mu+m_\alpha(\sigma)\right]\psi_{i\alpha}\right\}.
 \label{sm:vec:action}
\end{equation}
Here $a=1,\ldots,n_b$ and $\alpha=1,\ldots,n_D$ label the scalar and fermion species, respectively. The corresponding mass parameters are $m_a(\sigma)$ and $m_\alpha(\sigma)$, and $U(\sigma)$ is the renormalized potential for the auxiliary fields; temperature-independent local counterterms are understood. The Euclidean gamma matrices satisfy $\{\gamma^\mu,\gamma^\nu\}=2\delta^{\mu\nu}$. We work in a homogeneous equilibrium phase with $m_a^2\ge0$ and real $m_\alpha$.

At fixed $\sigma$, the matter fields in Eq.~\eqref{sm:vec:action} are free. Integrating them out gives the free-energy density $\mathcal F(T,\sigma)$ at leading order in $N$, where $\sigma$ denotes the dimensionful auxiliary fields collectively. Their equilibrium values $\sigma_*(T)$ obey the saddle-point equations $\partial_\sigma\mathcal F=0$. The chain rule therefore gives
\begin{equation}
 \frac{\dd\mathcal F(T,\sigma_*(T))}{\dd T}
 =\partial_T\mathcal F+\partial_\sigma\mathcal F\cdot\frac{\dd\sigma_*}{\dd T}
 =\partial_T\mathcal F,
 \label{sm:vec:stationarity}
\end{equation}
where all partial derivatives are evaluated at $\sigma=\sigma_*(T)$ and $\partial_T$ is taken at fixed $\sigma$. The term involving $\dd\sigma_*/\dd T$ vanishes by stationarity, even when the equilibrium auxiliary fields depend on temperature. Since $U$ and the zero-point energies have no explicit temperature dependence at fixed $\sigma$, only the thermal parts of the matter determinants contribute to the entropy density~\cite{DeWolfeRomatschke}:
\begin{equation}
 s=N\left[\sum_a s_B(T,m_a)+\sum_\alpha s_D(T,m_\alpha)\right]+O(N^0),
 \label{sm:vec:entropy}
\end{equation}
where $s_B$ and $s_D$ are the entropy densities of a free real scalar and a free two-component Dirac fermion, respectively, and all masses are evaluated at the equilibrium saddle. A Majorana fermion contributes one half of $s_D$.

Each entropy density is bounded by its massless value:
\begin{equation}
 s_B(T,m)\le s_B(T,0),\qquad s_D(T,m)\le s_D(T,0)\qquad(m^2\ge0).
 \label{sm:vec:massbound}
\end{equation}
To see this, consider a mode of mass $m$ and spatial momentum $\bm p$, and write $y=\sqrt{\bm p^2+m^2}/T$. The entropy densities are
\begin{equation}
 s_B(T,m)=\int\frac{\dd^2\bm p}{(2\pi)^2}\,h_B(y),\qquad s_D(T,m)=2\int\frac{\dd^2\bm p}{(2\pi)^2}\,h_F(y),
 \label{sm:vec:entropyintegrals}
\end{equation}
where $h_B(y)$ and $h_F(y)$ are the entropies of a bosonic mode and a fermionic mode, respectively. The factor of two accounts for the particle and antiparticle modes of the Dirac field. Explicitly,
\begin{equation}
 h_{B,F}(y)=\frac{y}{e^y\mp1}\mp\log(1\mp e^{-y}),\qquad h'_{B,F}(y)=-\frac{y e^y}{(e^y\mp1)^2}\le0,
 \label{sm:vec:modeentropy}
\end{equation}
where the upper and lower signs refer to bosons and fermions, respectively. Increasing $|m|$ increases $y$ for every momentum, so Eq.~\eqref{sm:vec:modeentropy} implies Eq.~\eqref{sm:vec:massbound}. Denoting the thermal coefficient of one massless Dirac fermion by $f_D$, we have
\begin{equation}
 \begin{aligned}
 s_B(T,0)&=3\fs T^2,& \fs&=\frac{\zeta(3)}{2\pi},\\
 s_D(T,0)&=3f_D T^2,& f_D&=\frac32\fs.
 \end{aligned}
 \label{sm:vec:massless}
\end{equation}
To relate the entropy bound to $\fv$, we now show that the stress-tensor three-point function agrees with that of the same free massless fields at leading order in $N$. Integrating out the matter fields gives an effective action for the auxiliary fields proportional to $N$. Vertices generated by matter loops and by the local interactions $NU(\sigma)$ are therefore $O(N)$, while an auxiliary-field propagator is $O(1/N)$. A connected tree has one fewer propagator than vertices and thus contributes at order $N$; a diagram containing a loop of auxiliary propagators is at most $O(N^0)$. The possible leading corrections to the three-point function of the free fields are therefore trees joining several vertices through auxiliary propagators. Every such tree has at least two end vertices, each connected to the rest by a single auxiliary propagator. The saddle-point equations cancel end vertices with no stress-tensor insertion, so each remaining end vertex must contain at least one stress tensor. Since there are only three stress-tensor insertions in total, at least one end vertex contains exactly one. This vertex is the two-point function of the stress tensor and the scalar bilinear coupled to the auxiliary field, which vanishes by conformal invariance~\cite{AGY,GurAriYacoby}. Thus these trees do not modify the three-point function at separated points.

The leading contribution is consequently a single free massless matter loop with three stress-tensor insertions, summed over $Nn_b$ real scalars and $Nn_D$ Dirac fermions. Weighting these contributions by $\fs$ and $f_D$, respectively, according to Eq.~\eqref{eq:SMvacdef}, we obtain
\begin{equation}
 \fv=N(n_b\fs+n_Df_D)+O(N^0).
 \label{sm:vec:fvac}
\end{equation}
Combining Eqs.~\eqref{sm:vec:entropy}, \eqref{sm:vec:massbound}, and~\eqref{sm:vec:massless} with $s=3fT^2$, we obtain
\begin{equation}
 f\le\fv+O(N^0).
 \label{sm:vec:thermalbound}
\end{equation}
Thus $f\le\fv$ at leading order in $N$.

The critical $O(N)$ model has $f/\fv=4/5$ at this order~\cite{Sachdev,Romatschke}. At the Gross--Neveu fixed point, the thermal mass of the fermions vanishes, giving $f=\fv$~\cite[Secs.~7.6--7.8]{Moshe}. The same argument applies to mixed scalar--fermion models~\cite{DeWolfeRomatschke} and sextic scalar theories in their stable conformal regime~\cite{Grable,Kvedaraite}.

For the critical $O(N)$ model, $F=N\Fs+O(N^0)$ at leading order, so $f/F=(4/5)\fs/\Fs+O(1/N)=2.39865\ldots+O(1/N)$. For the Gross--Neveu model with $N$ Dirac fermions, $F=NF_D+O(N^0)$ and $f/F=f_D/F_D+O(1/N)$. These comparisons concern the same fixed points as the stress-tensor bound, but their subleading corrections are independent data.

\subsection{Dimensional continuation near four dimensions}
For the Wilson--Fisher $O(N)$ model, we use $V=\mu^\epsilon\lambda(\phi_i\phi_i)^2/4!$ in $d=4-\epsilon$. The fixed-point coupling is $\lambda_*=48\pi^2\epsilon/(N+8)+O(\epsilon^2)$. Wick contractions evaluated at $d=4$ give $\delta f=-\lambda_*N(N+2)/3456$, and hence
\begin{equation}
 \frac{f_{\rm WF}(d)}{N\fs(d)}
 =1-\frac54\frac{N+2}{N+8}\epsilon+O(\epsilon^{3/2}),
 \label{eq:SMWFthermal}
\end{equation}
where $\fs(d)=\Gamma(d/2)\zeta(d)/\pi^{d/2}$ is the thermal coefficient of a free real scalar in $d$ dimensions.
The $O(\epsilon^{3/2})$ term is the leading thermal screening correction. With $\widetilde F(d):=\sin(\pi d/2)\log Z(\mathbb{S}^d)$, the interaction correction satisfies $\widetilde F_{\rm WF}-N\widetilde F_{\mathrm{s}}=O(\epsilon^3)$~\cite{FeiGiombiKlebanovTarnopolsky}, so
\begin{equation}
 \frac{f_{\rm WF}/\widetilde F_{\rm WF}}{\fs/\widetilde F_{\mathrm{s}}}
 =1-\frac54\frac{N+2}{N+8}\epsilon+O(\epsilon^{3/2}).
 \label{eq:SMWFentanglement}
\end{equation}
The free-field quantities $\fs$ and $\widetilde F_{\mathrm{s}}$ are evaluated in the same dimension $d$ as the interacting theory. This calculation provides a perturbative check by dimensional continuation. It does not describe an interacting four-dimensional CFT or establish a finite-$q$ R\'enyi inequality. A calculation in the presence of a conical defect requires separate renormalization, and its result cannot be inferred from the flat-space thermal coefficient.

\section{Chern--Simons vector theories at finite coupling}
\label{sm:cs:section}

Consider three-dimensional $U(N)$ Chern--Simons theory with a fixed number of fundamental flavors at leading planar order, and let $\lambda=N/\kappa$ be the 't Hooft coupling, where $\kappa$ is the renormalized level. We take the thermodynamic limit before the planar limit and work in a homogeneous equilibrium phase with nonnegative pole masses. To establish $f\le\fv$, we bound the entropy by a linear combination of free-field entropies with the same coefficients as the stress-tensor three-point function.

\subsection{Entropy from the matter determinants}
For one flavor in the fundamental representation, the Euclidean actions of the regular bosonic and fermionic theories on $\mathbb{S}^1_\beta\times\mathbb{R}^2$ are~\cite{AGGMY}
\begin{equation}
 \begin{aligned}
 I_B&=I_{\mathrm{CS}}+\int\dd^3x\,\left[(D_\mu\phi)^\dagger D_\mu\phi+\frac{\lambda_6}{6N^2}(\phi^\dagger\phi)^3\right],\\
 I_F&=I_{\mathrm{CS}}+\int\dd^3x\,\bar\psi\gamma^\mu D_\mu\psi,\\
 I_{\mathrm{CS}}&=\frac{i\kappa}{4\pi}\int\dd^3x\,\epsilon^{\mu\nu\rho}\Tr\left(A_\mu\partial_\nu A_\rho-\frac{2i}{3}A_\mu A_\nu A_\rho\right).
 \end{aligned}
 \label{sm:cs:action}
\end{equation}
Here $\phi$ is a complex scalar and $\psi$ a two-component Dirac fermion, each with $N$ color components. We use a Hermitian gauge field $A_\mu$ and $D_\mu=\partial_\mu-iA_\mu$; $\lambda_6$ is the scalar sextic coupling, and the trace is over color indices. The critical theories are obtained by introducing a scalar auxiliary field coupled to $\phi^\dagger\phi$ or $\bar\psi\psi$. A cubic interaction of the auxiliary field is also allowed in the fermionic theory.

The thermal holonomy has eigenvalues $e^{i\alpha_i}$, with phases $\alpha_i\in[-\pi,\pi]$. Their normalized density is $\rho(\alpha)=N^{-1}\sum_{i=1}^N\delta(\alpha-\alpha_i)$, understood in the large-$N$ limit. In a gauge with $A_\tau=T\operatorname{diag}(\alpha_1,\ldots,\alpha_N)$, a mode with holonomy phase $\alpha$ has Matsubara frequencies
\begin{equation}
 \omega_{n,B}=T(2\pi n-\alpha),\qquad \omega_{n,F}=T\bigl[2\pi(n+\tfrac12)-\alpha\bigr],\qquad n\in\mathbb Z.
 \label{sm:cs:frequencies}
\end{equation}
The planar calculation gives a thermal contribution of the same form as for free massive matter in the holonomy background~\cite{AGGMY,Dey}. For spatial momentum $\bm p$, the mode energy is $E=\sqrt{\bm p^2+m^2}$, where $m$ equals the pole mass at equilibrium. We denote $m$ and any additional dimensionful auxiliary fields collectively by $\sigma$. At fixed $\sigma$ and $\rho$, the Matsubara sums give the free-energy density
\begin{equation}
 \mathcal F(T,\sigma,\rho)=\mathcal F_0(\sigma)-2NT\int\frac{\dd^2\bm p}{(2\pi)^2}\int_{-\pi}^{\pi}\dd\alpha\,\rho(\alpha)L_{B,F}(E/T,\alpha),
 \label{sm:cs:free-energy}
\end{equation}
where $\mathcal F_0$ contains the vacuum and interaction terms and has no explicit temperature dependence at fixed $\sigma$. With $y=E/T$ and $r=e^{-y}$, the functions $L_B$ and $L_F$ are
\begin{equation}
 L_B(y,\alpha)=-\frac12\log(1-2r\cos\alpha+r^2),\qquad L_F(y,\alpha)=\frac12\log(1+2r\cos\alpha+r^2).
 \label{sm:cs:thermal-logs}
\end{equation}
The logarithms combine the particle and antiparticle factors $1\mp e^{-y+i\alpha}$ and $1\mp e^{-y-i\alpha}$, with the upper sign for bosons and the lower sign for fermions. Interactions enter through $\mathcal F_0$ and the equilibrium values of $\sigma$ and $\rho$.

As in Eq.~\eqref{sm:vec:stationarity}, the saddle-point equations eliminate terms proportional to the temperature derivatives of $\sigma$. The same argument applies to the equilibrium distribution $\rho$, with variations restricted by the temperature-independent constraints in Eq.~\eqref{sm:cs:cap}. We can therefore compute the entropy by differentiating Eq.~\eqref{sm:cs:free-energy} at fixed $\sigma$ and $\rho$. At fixed $E$ and $\alpha$, differentiating the thermal contribution defines the functions $h_{B,F}$ that enter the entropy:
\begin{equation}
 \frac{\partial}{\partial T}\left[T L_{B,F}(E/T,\alpha)\right]=L_{B,F}(y,\alpha)-y\partial_yL_{B,F}(y,\alpha)\equiv h_{B,F}(y,\alpha).
 \label{sm:cs:mode-entropy}
\end{equation}
Summing the contributions of all modes gives the entropy density
\begin{equation}
 s=-\left.\partial_T\mathcal F\right|_{\sigma,\rho}=2N\int\frac{\dd^2\bm p}{(2\pi)^2}\int_{-\pi}^{\pi}\dd\alpha\,\rho(\alpha)h_{B,F}(E/T,\alpha).
 \label{sm:cs:entropy}
\end{equation}
The factor $N$ counts colors. The factor of two compensates for the $1/2$ in $L_{B,F}$, which averages the particle and antiparticle contributions.

\subsection{Maximizing the holonomy average}
We first obtain an upper bound that does not require solving for the thermal holonomy. Set $\theta=\pi|\lambda|$, with $0<\theta<\pi$; the endpoints are obtained by continuity. The eigenvalue density satisfies~\cite{JainHolonomy}
\begin{equation}
 0\leq\rho(\alpha)\leq\frac1{2\theta},
 \qquad \int_{-\pi}^{\pi}\rho(\alpha)\,\dd\alpha=1.
 \label{sm:cs:cap}
\end{equation}
The entropy integrand is largest near $\alpha=0$, since
\begin{equation}
 \frac{\partial h_{B,F}}{\partial(\cos\alpha)}
 =\frac rD+\frac{yr(1-r^2)}{D^2}>0,
 \qquad D=1\mp2r\cos\alpha+r^2,
 \label{sm:cs:angular-monotonicity}
\end{equation}
where the upper sign applies to bosons. The average is therefore maximized by saturating the density bound closest to $\alpha=0$: $\rho_\theta=1/(2\theta)$ on $[-\theta,\theta]$ and zero outside this interval. To verify this, fix $y$ and use the pointwise inequality
\begin{equation}
 [\rho(\alpha)-\rho_\theta(\alpha)][h_{B,F}(y,\alpha)-h_{B,F}(y,\theta)]\leq0.
 \label{sm:cs:pointwise}
\end{equation}
Integrating this inequality proves the stated maximum: the term proportional to $h_{B,F}(y,\theta)$ cancels because both distributions are normalized.

\subsection{Comparison with free-field entropies}
We next bound this maximal holonomy average by massless free-field entropies. The averages over $\rho_\theta$ admit the integral representation
\begin{equation}
 \begin{gathered}
 H(x,y)=\frac12\int_y^\infty\frac{u}{\cosh u-x}\,\dd u\qquad(-1\le x\le1),\\
 \overline h_B(y)=\frac{\sin\theta}{\theta}H(\cos\theta,y),
 \qquad \overline h_F(y)=\frac{\sin\theta}{\theta}H(-\cos\theta,y).
 \end{gathered}
 \label{sm:cs:kernel}
\end{equation}
This representation makes the effect of the pole mass explicit: $H(x,y)$ decreases with $y$, so replacing $E=\sqrt{\bm p^2+m^2}$ by $|\bm p|$ gives an upper bound. For example, the bosonic Fourier series gives
\begin{equation}
 \frac{\dd\overline h_B}{\dd y}
 =-\frac{y}{\theta}\sum_{k\ge1}e^{-ky}\sin(k\theta)
 =-\frac{y\sin\theta}{2\theta(\cosh y-\cos\theta)}.
\end{equation}
Together with $\overline h_B(\infty)=0$, this proves the integral representation; alternating signs give the fermionic result. Convexity in the first argument follows from
\begin{equation}
 \partial_x^2H(x,y)=\int_y^\infty\frac{u\,\dd u}{(\cosh u-x)^3}>0
 \qquad(-1<x<1,\ y>0).
\end{equation}
To compare the resulting massless contribution with free fields, note that $H(1,y)$ and $H(-1,y)$ are the entropies of free bosonic and fermionic modes. Convexity in the first argument then gives
\begin{equation}
 H(\cos\theta,y)\leq\frac{1+\cos\theta}{2}H(1,y)+\frac{1-\cos\theta}{2}H(-1,y).
 \label{sm:cs:entropy-comparison}
\end{equation}
At $y=|\bm p|/T$, this bounds the bosonic contribution by a linear combination of massless free-field entropies. For fermions, replacing $\cos\theta$ by $-\cos\theta$ exchanges the two weights.

It remains to identify the coefficients of this combination with those defining $\fv$. In the basis defined by one free real scalar and one free Majorana fermion, the coefficients of the stress-tensor three-point function for bosonic and fermionic matter are, respectively,~\cite{AGY,ParityCollider}
\begin{equation}
 (n_s,n_M)_B=N\frac{\sin\theta}{\theta}(1+\cos\theta,1-\cos\theta),
 \qquad (n_s,n_M)_F=N\frac{\sin\theta}{\theta}(1-\cos\theta,1+\cos\theta).
 \label{sm:cs:ttt}
\end{equation}
Multiplying the weights in Eq.~\eqref{sm:cs:entropy-comparison} and its fermionic counterpart by the factor $2N\sin\theta/\theta$ from Eqs.~\eqref{sm:cs:entropy} and~\eqref{sm:cs:kernel} gives exactly these coefficients. The mode entropies $H(1,y)$ and $H(-1,y)$ integrate to the free entropy densities $3\fs T^2$ and $3f_MT^2$, respectively, with $f_M=3\fs/4$ as in Sec.~\ref{sec:SMconventions}. Thus
\begin{equation}
 s(T)\leq3\fs\left(n_s+\frac34n_M\right)T^2=3\fv T^2,
 \qquad f\leq\fv.
 \label{sm:cs:bound}
\end{equation}
The bound on $f$ follows from $s=3fT^2$ at a conformal fixed point. The coefficients in Eq.~\eqref{sm:cs:ttt} are unchanged at this order by the Legendre transformation between the regular and critical theories. For $0<\theta<\pi$, strict convexity of $H$ in its first argument gives $f<\fv$. For several flavors, the bound follows by addition when the leading coefficients of the stress-tensor three-point function are additive.
\section{Gravitational bounds and higher-curvature tests}
\label{sm:grav:section}

\subsection{Einstein gravity with neutral scalars}

In two-derivative Einstein gravity coupled to neutral scalars, the entropy density at fixed temperature is bounded above by its Schwarzschild--AdS value. Related entropy bounds along holographic renormalization-group flows were proposed in Ref.~\cite{Wen}.

Consider the classical bulk action
\begin{equation}
 I=\frac1{16\pi G_{d+1}}\int\dd^{d+1}x\sqrt{-g}\left[\mathcal R-\frac12G_{IJ}(\phi)\partial_\mu\phi^I\partial^\mu\phi^J-V(\phi)\right],
 \label{sm:grav:action}
\end{equation}
where $G_{d+1}$ is Newton's constant, $\mathcal R$ is the Ricci scalar, $G_{IJ}$ is the positive-definite scalar kinetic matrix, and $V$ is the scalar potential. The CFT vacuum is dual to AdS with radius $L$. We consider thermal equilibrium described by a regular, static, homogeneous and isotropic planar black brane with radial scalar profiles, vanishing sources, and conformal boundary conditions. With radial coordinate $r$ and the horizon at $r=r_H$, write the metric as
\begin{equation}
 \dd s^2=e^{2A(r)}[-h(r)\dd t^2+\dd\bm x^2]+\frac{\dd r^2}{h(r)},\qquad h(r_H)=0,\quad h(\infty)=1,\quad A'(\infty)=L^{-1}.
 \label{sm:grav:metric}
\end{equation}
Here $A(r)$ determines the warp factor $e^{2A(r)}$, and $h(r)$ is the blackening function, which vanishes at the horizon. Primes denote derivatives with respect to $r$, and we write $A_H=A(r_H)$ and $h'_H=h'(r_H)$. The relevant Einstein equations are~\cite{GubserNellore}
\begin{equation}
 A''=-\frac{G_{IJ}\phi^{I\prime}\phi^{J\prime}}{2(d-1)}\le0,\qquad h''+dA'h'=0.
 \label{sm:grav:equations}
\end{equation}
The inequality $A''\le0$ follows from the null energy condition, which is satisfied because $G_{IJ}$ is positive definite. Together with $A'(\infty)=L^{-1}$, it gives $A'\ge L^{-1}$ and hence $A(r)-A_H\ge(r-r_H)/L$. Integrating the second equation with $h(r_H)=0$ and $h(\infty)=1$ then gives
\begin{equation}
 h'_H=\left[\int_{r_H}^{\infty}e^{-d[A(r)-A_H]}\,\dd r\right]^{-1}\ge\frac dL.
 \label{sm:grav:horizonidentity}
\end{equation}
The Hawking temperature and Bekenstein--Hawking entropy density are
\begin{equation}
 T=\frac{e^{A_H}h'_H}{4\pi},\qquad s=\frac{e^{(d-1)A_H}}{4G_{d+1}}.
 \label{sm:grav:thermodynamics}
\end{equation}
At fixed temperature, the lower bound $h'_H\ge d/L$ therefore bounds $e^{A_H}$, and hence the entropy density, from above. Denoting the thermal coefficient of the planar Schwarzschild--AdS solution with the same $L$ and $G_{d+1}$ by $f_{\rm E}$~\cite{KovtunRitz}, we obtain
\begin{equation}
 \frac{s}{d f_{\rm E}T^{d-1}}
 =\left(\frac{d}{Lh'_H}\right)^{d-1}\le1,
 \qquad f_{\rm E}=\frac1{4dG_{d+1}}\left(\frac{4\pi L}{d}\right)^{d-1}.
 \label{sm:grav:deficit}
\end{equation}
Equality requires constant scalars throughout the exterior. Conformal thermodynamics gives $s=dfT^{d-1}$ and hence $f\le f_{\rm E}$. The bound is independent of the scalar potential away from the AdS vacuum.

To express $f_{\rm E}$ and $\fv$ in terms of CFT data, we use the holographic relations in classical Einstein gravity~\cite{Henningson,Hofman,Bueno,KovtunRitz}
\begin{align}
 d=4:&\qquad a=c=\frac{\pi L^3}{8G_5},\qquad t_2=t_4=0,\notag\\
 d=3:&\qquad \ct=\frac{3L^2}{\pi^3G_4},\qquad F=\frac{\pi L^2}{2G_4},\qquad t_4=0.
 \label{sm:grav:cftdata}
\end{align}
Substituting these data into Eqs.~\eqref{eq:SMcolliderfour}, \eqref{eq:SMcolliderthree}, and~\eqref{sm:grav:deficit} gives
\begin{align}
 d=4:&\qquad f_{\rm E}=\frac{\pi^2a}{2},\qquad \fv=\frac{2\pi^2a}{3},\qquad \frac f\fv\le\frac34,
 \label{sm:grav:four}\\
 d=3:&\qquad f_{\rm E}=\frac{4\pi^2L^2}{27G_4},\qquad \fv=\frac{14\zeta(3)L^2}{\pi^2G_4},\qquad \frac f\fv\le\frac{2\pi^4}{189\zeta(3)}=0.8575168\ldots.
 \label{sm:grav:three}
\end{align}
In three dimensions, $f/F\le8\pi/27<\fs/\Fs$ follows directly, without the BCLM conjecture.
\subsection{Higher-curvature gravitational tests}
\label{sec:SMhighercurvature}

The preceding argument assumes a two-derivative Einstein action. We now test the thermal inequalities in explicit higher-curvature models. These are algebraic checks of the specified gravitational models and branches; energy-flux positivity alone does not establish their stability or a UV-complete unitary CFT dual.

For five-dimensional Gauss--Bonnet gravity, let $u=L_0^2/L^2$ satisfy $1-u+\lambda_{\rm GB}u^2=0$ on the branch connected to Einstein gravity, where $L_0$ is the bare radius and $L$ the effective AdS radius. With $m=\lambda_{\rm GB}u$ and $K=\pi L^3/(8G_5)$, the anomaly and planar black-brane data are~\cite{HMSY,MyersPaulosSinha}
\begin{equation}
 a=K(1-6m),\qquad c=K(1-2m),\qquad
 f=\frac{\pi^2K}{2}u^3,\qquad
 f_{\rm vac}=\frac{2\pi^2K}{3}(1+2m).
 \label{eq:SMGBdata}
\end{equation}
The factor $u^3$ includes the normalization of boundary time. In terms of $x=c/a$, the energy-flux constraints give $2/3\le x\le2$, and
\begin{equation}
 \frac{f}{4\pi^2a}=\frac{(3x-1)^4}{2(5x-1)^3},\qquad
 \frac{f}{f_{\rm vac}}=\frac{3(3x-1)^4}{(5x-1)^3(2x-1)}.
 \label{eq:SMGBratios}
\end{equation}
The first function increases throughout this interval. The derivative of the second has the sign of $7x-5$, so its maximum is also at an endpoint. Comparing the endpoints gives $f/(4\pi^2a)\le625/1458\simeq0.428669$ and $f/f_{\rm vac}\le625/729\simeq0.857339$. These are bounds for the planar branch. Some energy-flux-allowed couplings have known plasma instabilities~\cite{MyersPaulosSinha}; such points must not be counted as verified equilibrium CFT examples.

The cubic quasi-topological model supplies a test with nonzero $t_4$. Use the normalization of Ref.~\cite{MyersPaulosSinha} and define
\begin{equation}
 1-u+\lambda_{\rm GB}u^2+\mu u^3=0,\qquad
 m=\lambda_{\rm GB}u,\qquad n=\mu u^2,\qquad u=(1-m-n)^{-1}.
 \label{eq:SMQTGvariables}
\end{equation}
The variable $n$ in this subsection is a gravitational coupling combination, not a field multiplicity or a R\'enyi index. Writing $A=1-6m+9n$ and $C=1-2m-3n$, the relevant data are
\begin{equation}
 a=KA,\quad c=KC,\quad
 t_2=\frac{24(m-87n)}C,\quad t_4=\frac{3780n}C,\quad
 f=\frac{\pi^2K}{2}u^3,\quad
 f_{\rm vac}=\frac{2\pi^2K}{3}(1+2m+153n).
 \label{eq:SMQTGdata}
\end{equation}
The last equality follows directly by substituting $c,t_2,t_4$ into Eq.~\eqref{eq:SMcolliderfour}. Positivity of energy flux requires
\begin{equation}
 1-10m+189n\ge0,\qquad 1+2m-855n\ge0,\qquad
 1+6m+1317n\ge0.
 \label{eq:SMQTGtriangle}
\end{equation}
These inequalities delimit the triangle in Table~\ref{tab:SMQTG}. Its vertices have the $\langle TTT\rangle$ structures of a free scalar, Weyl fermion, or Maxwell field, but the gravitational thermodynamics is not that of a free theory.

\begin{table}[htbp]
\centering
\renewcommand{\arraystretch}{1.2}
\begin{tabular}{lccccc}
\toprule
Tensor-structure corner & $m$ & $n$ & $c/a$ & $f/(4\pi^2a)$ & $f/f_{\rm vac}$\\
\midrule
Scalar & $29/227$ & $1/681$ & $3$ & $0.7674058382$ & $0.7674058382$\\
Weyl & $47/596$ & $-1/894$ & $18/11$ & $0.3083515478$ & $0.9691048644$\\
Maxwell & $-181/647$ & $1/1941$ & $18/31$ & $0.0222541505$ & $0.6898786648$\\
\bottomrule
\end{tabular}
\caption{Formal planar-branch data at the energy-flux vertices of cubic quasi-topological gravity. The ratios follow from Eq.~\eqref{eq:SMQTGdata}; the labels specify vacuum tensor structures, not known free-field duals.}
\label{tab:SMQTG}
\end{table}

The vertex checks extend analytically to the entire triangle. On it, $A$, $1+2m+153n$, and $1-m-n$ are positive. The function $g(m,n)=(1-m-n)^{-3}$ is convex. If a point is a convex combination of vertices $v_i$ with weights $w_i$, then for either positive linear denominator $D_*(m,n)=8A$ or $D_*(m,n)=4(1+2m+153n)/3$,
\begin{equation}
 \frac{g(\sum_iw_iv_i)}{D_*(\sum_iw_iv_i)}
 \le\frac{\sum_iw_i g(v_i)}{\sum_iw_iD_*(v_i)}
 \le\max_i\frac{g(v_i)}{D_*(v_i)}.
 \label{eq:SMQTGconvexproof}
\end{equation}
Thus $f/(4\pi^2a)\le0.767406$ and $f/f_{\rm vac}\le0.969105$ are rigorous algebraic bounds on this branch throughout the energy-flux triangle. The formal Maxwell corner also has $f/(4\pi^2a)<1/31$, below the free-Maxwell value. Because stability and UV completion impose additional requirements~\cite{MyersPaulosSinha}, this observation by itself is not a counterexample in an established unitary CFT. No claim about a full R\'enyi curve follows solely from these flat-space endpoint checks.

\section{Spherical R\'enyi entropy and the thermal limit}
\label{sm:renyi:section}

Let $\rho_B$ be the reduced density matrix for a ball of radius $R$ in the vacuum. The R\'enyi entropy with index $q>0$ and its universal contribution are
\begin{equation}
 \begin{gathered}
 S_q=\frac{\log\operatorname{Tr}\rho_B^q}{1-q},\\
 S_q^{\rm univ}=
 \begin{cases}
 -4a_q\log(R/\epsilon),&d=4,\\
 -F_q,&d=3,
 \end{cases}
 \qquad a_1=a,\quad F_1=F.
 \end{gathered}
 \label{sm:renyi:definition}
\end{equation}
Here $\epsilon$ is an ultraviolet cutoff, and the limit $q\to1$ gives the entanglement entropy. The conformal map relates $\rho_B$ to a thermal state at $T_0=1/(2\pi R)$ on hyperbolic space $\mathbb H^{d-1}_R$ with curvature radius $R$~\cite{CHM}.

Let $s_H(T)$ denote the entropy density on hyperbolic space, and write $t=T/T_0$. For the free scalar and the CFTs dual to Einstein gravity considered below, the representation in terms of thermal entropy~\cite{HMSY} gives the normalized R\'enyi coefficients
\begin{equation}
 \begin{cases}
 a_q/a,&d=4,\\
 F_q/F,&d=3,
 \end{cases}
 =\frac{q}{1-q}\int_1^{1/q}\frac{s_H(T_0t)}{s_H(T_0)}\,\dd t,
 \qquad 0<q<1.
 \label{sm:renyi:entropyintegrals}
\end{equation}
The regulated hyperbolic volume cancels in these ratios. Since the integration weight is positive, an upper bound on the normalized entropy density gives an upper bound on the normalized R\'enyi coefficient.

\subsection{Thermal limits and comparison with a free scalar}

As $q\to0$, the temperature is large compared with the curvature scale. The leading high-temperature term $dfT^{d-1}$ in the entropy density contributes $f\,\operatorname{Vol}(\mathbb H^{d-1}_R)/(2\pi Rq)^{d-1}$ to $S_q$. The universal contributions to the regulated hyperbolic volume are~\cite{HMSY}
\begin{equation}
 \left.\operatorname{Vol}(\mathbb H^{d-1}_R)\right|_{\rm univ}=
 \begin{cases}
 -2\pi R^3\log(R/\epsilon),&d=4,\\
 -2\pi R^2,&d=3.
 \end{cases}
 \label{sm:renyi:hyperbolic-volume}
\end{equation}
Using the definitions of $a_q$ and $F_q$ in Eq.~\eqref{sm:renyi:definition} therefore gives the thermal limits~\cite{Kusuki}
\begin{equation}
 \lim_{q\to0}16\pi^2q^3a_q=f\quad(d=4),
 \qquad
 \lim_{q\to0}2\pi q^2F_q=f\quad(d=3).
 \label{sm:renyi:thermal-limits}
\end{equation}
Curvature corrections and terms localized on the entangling surface do not change these limits.

We compare the R\'enyi coefficients after normalizing by their values at $q=1$. For a free real scalar, denote the corresponding coefficients by $a_{q,\mathrm{s}}$ and $F_{q,\mathrm{s}}$, with $a_{1,\mathrm{s}}=\as$ and $F_{1,\mathrm{s}}=\Fs$. Define
\begin{equation}
 R_4(q)=\frac{a_q/a}{a_{q,\mathrm{s}}/\as},
 \qquad
 R_3(q)=\frac{F_q/F}{F_{q,\mathrm{s}}/\Fs}.
 \label{sm:renyi:normalized-ratios}
\end{equation}
Both ratios equal one at $q=1$, while their thermal limits are
\begin{equation}
 \lim_{q\to0}R_4(q)=\frac{f}{4\pi^2a},
 \qquad
 \lim_{q\to0}R_3(q)=\frac{f/F}{\fs/\Fs}.
 \label{sm:renyi:normalized-endpoints}
\end{equation}
In the limit $q\to0$, the stronger inequalities $R_d(q)\le1$ for $0<q\le1$ would imply the proposed bounds in Eq.~\eqref{eq:headline}: $f\le4\pi^2a$ in four dimensions and $f\le(\fs/\Fs)F$ in three dimensions.

The first derivatives of $a_q$ and $F_q$ with respect to $q$, evaluated at $q=1$, are fixed by the stress-tensor two-point function~\cite{PerlmutterRenyi}:
\begin{equation}
 a'_1=-\frac c2,
 \qquad
 F'_1=-\frac{\pi^4}{12}\ct.
 \label{sm:renyi:first-derivatives}
\end{equation}
With the normalization in Sec.~\ref{sec:SMconventions}, applying Eq.~\eqref{sm:renyi:first-derivatives} to the theory and the free scalar, for which $c=3a$ in four dimensions, gives
\begin{align}
 R'_4(1)&=\frac{3a-c}{2a},\label{sm:renyi:four-derivative}\\
 R'_3(1)&=\frac{\pi^4}{12}\left(\frac{\cts}{\Fs}-\frac{\ct}{F}\right).
 \label{sm:renyi:three-derivative}
\end{align}
In four dimensions, the conformal collider bounds imply $c\le3a$~\cite{Hofman,ColliderProof}; in three dimensions, the BCLM conjecture gives $\ct/F\le\cts/\Fs$~\cite{Bueno}. If the corresponding inequality is strict, then $R'_d(1)>0$, so $R_d(q)<1$ for $q<1$ sufficiently close to one.

\subsection{Free-field verification in four dimensions}
The spherical R\'enyi coefficients for free fields give an explicit test throughout $0<q\le1$~\cite{KlebanovRenyi,FursaevRenyi,DonnellyWall}:
\begin{equation}
 \frac{a_{q,\mathrm{s}}}{\as}=\frac{(1+q)(1+q^2)}{4q^3},\qquad
 \frac{a_{q,W}}{a_W}=\frac{(1+q)(7+37q^2)}{88q^3},\qquad
 \frac{a_{q,V}}{a_V}=\frac{1+q+31q^2+91q^3}{124q^3}.
 \label{eq:SMfreeRenyi}
\end{equation}
The Maxwell result includes the edge contribution in the geometric prescription that reproduces the conformal anomaly. In this convention
\begin{equation}
 a_{q,V}=\frac{(1+q)(1+31q^2)}{720q^3}+\frac1{12},
 \label{eq:SMMaxwellEdge}
\end{equation}
where the last term is independent of $q$ and restores $a_{1,V}=31/180$; it leaves the thermal limit unchanged. Subtracting the normalized coefficients gives
\begin{align}
 \frac{a_{q,\mathrm{s}}}{\as}-\frac{a_{q,W}}{a_W}
 &=\frac{15(1+q)(1-q^2)}{88q^3}\ge0,\\
 \frac{a_{q,\mathrm{s}}}{\as}-\frac{a_{q,V}}{a_V}
 &=\frac{15(1-q)(2q^2+2q+1)}{62q^3}\ge0.
\end{align}
For a direct product of free theories, $a_q/a$ is an $a$-weighted average of these ratios. The normalized R\'enyi coefficient is therefore bounded above by its free-scalar value for any such free theory.

\subsection{Verification in Einstein gravity}
\label{sm:renyi:gravity}

In classical Einstein gravity, assume that the homogeneous Schwarzschild--AdS black hole with a hyperbolic horizon computes the equilibrium partition function for every $T\ge T_0$. Let $x$ be its horizon radius in units of the AdS radius. Its temperature and entropy satisfy~\cite{HMSY}
\begin{equation}
 2t=dx-\frac{d-2}{x},\qquad
 \frac{s_H(T_0t)}{s_H(T_0)}=x^{d-1}.
 \label{sm:renyi:einstein-entropy}
\end{equation}
The temperature increases with $x$, and $x=1$ at $t=1$. Thus $t\ge1$ implies $x\ge1$, and
\begin{equation}
 2(t-x)=(d-2)(x-x^{-1})\ge0,
 \qquad
 \frac{s_H(T_0t)}{s_H(T_0)}\le t^{d-1}.
 \label{sm:renyi:einstein-entropy-bound}
\end{equation}
In four dimensions, inserting this bound into Eq.~\eqref{sm:renyi:entropyintegrals} gives
\begin{equation}
 \frac{a_q}{a}\le\frac{1+q+q^2+q^3}{4q^3}
 =\frac{a_{q,\mathrm{s}}}{\as},\qquad 0<q\le1.
 \label{sm:renyi:gravity-four-chain}
\end{equation}
The equality uses the known free-scalar R\'enyi coefficient~\cite{KlebanovRenyi}.

To complete the three-dimensional comparison, let $s_{H,\mathrm{s}}(T)$ denote the entropy density of a free real scalar on hyperbolic space. Its spectrum gives~\cite{KlebanovRenyi}
\begin{equation}
 \begin{aligned}
 \frac{s_{H,\mathrm{s}}(T)}{T^2}&=\frac1{2\pi}\int_0^\infty y\tanh(\pi RTy)h_B(y)\,\dd y,\\
 h_B(y)&=\frac{y}{e^y-1}-\log(1-e^{-y})>0,
 \end{aligned}
 \label{sm:renyi:scalarscaled}
\end{equation}
where $y=E/T$ and $h_B(y)$ is the entropy of a bosonic mode with energy $E$. Monotonicity of $\tanh$ implies
\begin{equation}
 \frac{s_{H,\mathrm{s}}(T_0t)}{s_{H,\mathrm{s}}(T_0)}\ge t^2,\qquad t\ge1.
 \label{sm:renyi:scalar-entropy-bound}
\end{equation}
Combining this lower bound with the entropy bound for Einstein gravity in Eq.~\eqref{sm:renyi:einstein-entropy-bound} and integrating as in Eq.~\eqref{sm:renyi:entropyintegrals} yields
\begin{equation}
 \frac{F_q}F\le\frac{1+q+q^2}{3q^2}\le\frac{F_{q,\mathrm{s}}}{\Fs},\qquad0<q\le1.
 \label{sm:renyi:gravity-three}
\end{equation}
This verifies the normalized R\'enyi comparison in three dimensions without using the BCLM conjecture.
\section{Numerical evidence at finite \texorpdfstring{$N$}{N}}\label{sec:SMnumerics}
\subsection{The Ising CFT}
\label{sec:SMIsing}

In the free-field basis of Eq.~\eqref{eq:SMvacdef}, Table~1 of Ref.~\cite{IsingTTT} gives
\begin{equation}
 n_s=0.933444559(75),\qquad n_M=0.013094116(33).
\end{equation}
These values give $\fv\simeq0.18045917$. The uncertainties in $n_s$ and $n_M$ are estimated by sampling the region allowed by the conformal bootstrap under the spectral assumptions of that analysis.

The Monte Carlo calculation in Table~II of Ref.~\cite{Bulgarelli} gives the critical Casimir amplitude $\Delta=-0.1531(4)$ for a film with periodic boundary conditions. For film thickness $\beta$ and transverse area $A$, the excess contribution to $-\log Z$ is $\Delta A/\beta^2$. Comparing with Eq.~\eqref{eq:SMfdef} gives $f=-\Delta$, so
\begin{equation}
 f=0.1531(4),\qquad \frac f{\fv}=0.8484(22).
\end{equation}
The independent fuzzy-sphere estimate $F=0.0612(5)$~\cite{Hu} gives
\begin{equation}
 \frac{f/F}{\fs/\Fs}=0.834(7).
\end{equation}

\subsection{The \texorpdfstring{$O(2)$}{O(2)} CFT}

The bootstrap analysis of mixed correlators of the conserved $U(1)$ current and the stress tensor in Ref.~\cite{O2TTT} holds $\ct/(2\cts)=0.944$ fixed and constrains the relative coefficient of the free-fermion structure in $\langle TTT\rangle$:
\begin{equation}
 x:=\frac{n_M}{n_s+n_M}=0.026(18).
 \label{sm:num:O2input}
\end{equation}
The interval was obtained with the spectral assumptions in Table~2 of that reference. We use the result obtained without an additional constraint on $\langle JJT\rangle$, where $J$ is the conserved $U(1)$ current.

Equation~\eqref{eq:SMcolliderthree} then gives
\begin{equation}
 \fv=\fs\frac{\ct}{\cts}\left(1-\frac x4\right)=0.35885(163).
 \label{sm:num:O2vac}
\end{equation}
The Monte Carlo result in Ref.~\cite{Bulgarelli}, $\Delta/N=-0.1520(8)$ with $N=2$, gives
\begin{equation}
 f=0.3040(16),\qquad \frac f{\fv}\simeq0.847.
\end{equation}
The quoted uncertainty in $\fv$ includes only the uncertainty in $x$. Earlier estimates of the critical Casimir amplitude differ by more than their quoted statistical errors, indicating residual systematic uncertainties.

\end{document}